\documentclass[aps,prd,preprint,nopreprintnumbers,nofootinbib,showpacs,tightenlines,superscriptaddress]{revtex4}

\usepackage{graphicx}
\usepackage{dcolumn}
\usepackage{bm}
\usepackage{amsmath}

\begin{document}

\preprint{CLNS 07/2014}       
\preprint{CLEO 07-18}         

\title{\boldmath Determination of the $D^0 \to K^+\pi^-$ Relative Strong Phase
Using Quantum-Correlated Measurements in $e^+e^- \to D^0 \bar D^0$ at CLEO}

\author{D.~M.~Asner}
\author{K.~W.~Edwards}
\author{P.~Naik}
\affiliation{Carleton University, Ottawa, Ontario, Canada K1S 5B6}
\author{R.~A.~Briere}
\author{T.~Ferguson}
\author{G.~Tatishvili}
\author{H.~Vogel}
\author{M.~E.~Watkins}
\affiliation{Carnegie Mellon University, Pittsburgh, Pennsylvania 15213, USA}
\author{J.~L.~Rosner}
\affiliation{Enrico Fermi Institute, University of
Chicago, Chicago, Illinois 60637, USA}
\author{J.~P.~Alexander}
\author{D.~G.~Cassel}
\author{J.~E.~Duboscq}
\author{R.~Ehrlich}
\author{L.~Fields}
\author{L.~Gibbons}
\author{R.~Gray}
\author{S.~W.~Gray}
\author{D.~L.~Hartill}
\author{B.~K.~Heltsley}
\author{D.~Hertz}
\author{C.~D.~Jones}
\author{J.~Kandaswamy}
\author{D.~L.~Kreinick}
\author{V.~E.~Kuznetsov}
\author{H.~Mahlke-Kr\"uger}
\author{D.~Mohapatra}
\author{P.~U.~E.~Onyisi}
\author{J.~R.~Patterson}
\author{D.~Peterson}
\author{D.~Riley}
\author{A.~Ryd}
\author{A.~J.~Sadoff}
\author{X.~Shi}
\author{S.~Stroiney}
\author{W.~M.~Sun}
\author{T.~Wilksen}
\affiliation{Cornell University, Ithaca, New York 14853, USA}
% ...
\collaboration{CLEO Collaboration} 
\noaffiliation

\date{February 15, 2008}

\begin{abstract} 
We exploit the quantum coherence between pair-produced $D^0$ and
$\bar D^0$ in $\psi(3770)$ decays to study charm mixing, which is characterized
by the parameters $x$ and $y$, and to make a first
determination of the relative strong phase $\delta$ between
doubly Cabibbo-suppressed $D^0\to K^+\pi^-$
and Cabibbo-favored $\bar D^0\to K^+\pi^-$.
We analyze a sample of 1.0 million $D^0\bar D^0$
pairs from 281 ${\rm pb}^{-1}$ of $e^+e^-$ collision data collected with the
CLEO\nobreakdash-c detector at $E_{\rm cm}=3.77$ GeV.
By combining CLEO\nobreakdash-c measurements
with branching fraction input and time-integrated measurements of 
$R_{\rm M}\equiv (x^2+y^2)/2$ and 
$R_{\rm WS}\equiv\Gamma(D^0\to K^+\pi^-)/\Gamma(\bar D^0\to K^+\pi^-)$
from other experiments, we find
$\cos\delta = 1.03^{+0.31}_{-0.17}\pm 0.06$,
where the uncertainties are statistical and systematic, respectively.
In addition, by further including external measurements of charm mixing
parameters, we obtain an alternate
measurement of $\cos\delta = 1.10\pm 0.35\pm 0.07$, as well as
$x\sin\delta = (4.4^{+2.7}_{-1.8}\pm 2.9)\times 10^{-3}$
and $\delta = (22^{+11}_{-12}$$^{+9}_{-11})^\circ$.
\end{abstract}

\maketitle

\section{Introduction}

In the Standard Model, $D^0$-$\bar D^0$ mixing is suppressed both by the
Glashow-Iliopoulos-Maiani mechanism~\cite{gim}
and by Cabibbo-Kobayashi-Maskawa matrix elements~\cite{ckm},
although sizeable mixing could arise
from new physics~\cite{Bianco:2003vb}.
Charm mixing is conventionally described by two small dimensionless parameters:
\begin{eqnarray}
x &=& 2\frac{M_2 - M_1}{\Gamma_2 + \Gamma_1} \\
y &=& \frac{\Gamma_2 - \Gamma_1}{\Gamma_2 + \Gamma_1},
\end{eqnarray}
where $M_{1,2}$ and $\Gamma_{1,2}$ are the masses and widths, respectively,
of the neutral $D$ meson $CP$ eigenstates, $D_1$ ($CP$-odd) and
$D_2$ ($CP$-even), which are defined as follows:
\begin{eqnarray}
|D_1\rangle \equiv \frac{|D^0\rangle + |\bar D^0\rangle}{\sqrt{2}} \\
|D_2\rangle \equiv \frac{|D^0\rangle - |\bar D^0\rangle}{\sqrt{2}},
\end{eqnarray}
assuming $CP$ conservation.  The mixing probability is then denoted by
$R_{\rm M}\equiv (x^2+y^2)/2$, and the width of the $D^0$ and $\bar D^0$
flavor eigenstates is $\Gamma\equiv (\Gamma_1+\Gamma_2)/2$.

By focusing on $D^0$ decay times, recent experiments have made
precise measurements of $D^0$-$\bar D^0$ mixing
parameters~\cite{kpiBelle,ycpBelle,kspipiBelle,kpiBABAR,kpiCDFnew}
that highlight the need for information on the relative phase between
the Cabibbo-favored decay $D^0 \to K^-\pi^+$ and the
doubly Cabibbo-suppressed (DCS) decay $\bar D^0 \to K^-\pi^+$.
Direct measurements of $y$ come from comparing decay times in 
$D^0\to K^-\pi^+$ to those in
$D^0$ transitions to the $CP$-even eigenstates $K^+K^-$ and $\pi^+\pi^-$.
Time-dependent studies of the resonant substructure in $D^0\to K^0_S\pi^+\pi^-$
provide $x$ as well as $y$.
In contrast, an indirect measure of $y$ is provided by the
``wrong-sign'' process $D^0\to K^+\pi^-$, where
interference between the DCS amplitude and the
mixing amplitude manifests itself in the $D^0$ decay time distributions.
These analyses are sensitive to $y'\equiv y\cos\delta - x\sin\delta$, where
$-\delta$ is the relative phase
between the DCS amplitude and the corresponding Cabibbo-favored
$\bar D^0\to K^+\pi^-$ amplitude:
$\langle K^+\pi^-|D^0\rangle / \langle K^+\pi^-|\bar D^0\rangle\equiv r e^{-i\delta}$.  We adopt a convention in which $\delta$ corresponds to a strong
phase, which vanishes in the SU(3) limit~\cite{Gronau:2001nr}.
The magnitude $r$ of the amplitude ratio is approximately 0.06.

Measurements of $y$ and $y'$ have both attained a precision of
${\cal O}(10^{-3})$.  However, because $\delta$ has not previously been
measured,
these separate determinations of $y$ and $y'$ have not been directly
comparable.  Thus, even a modest measurement of $\delta$ can significantly
improve the overall knowledge of charm mixing parameters.

In this article, we present a first determination of $\delta$ that
takes advantage of the correlated production of $D^0$ and $\bar D^0$ mesons 
in $e^+e^-$ collisions produced at the Cornell Electron Storage Ring and
collected with the CLEO-c detector.
If a collision produces no accompanying particles, the $D^0\bar D^0$ pair is
in a
quantum-coherent $C=-1$ state.  Because the initial state (the virtual photon)
has $J^{PC} = 1^{--}$, there follows a set of selection rules for the decays
of the $D^0$ and
$\bar D^0$~\cite{Kingsley:1975fe,Okun:1975di,Kingsley:1976cd,Goldhaber:1976fp,Bigi:1986dp,Bigi:1986rj,Bigi:1989ah,Xing:1996pn,Gronau:2001nr,Bianco:2003vb,Atwood:2002ak}.
For example, both $D^0$ and $\bar D^0$
cannot decay to $CP$ eigenstates with the same eigenvalue.  On the other hand,
decays to $CP$ eigenstates of opposite eigenvalue are enhanced by a factor of
two.
More generally, final states that can be reached by both $D^0$ and $\bar D^0$
(such as $K^-\pi^+$) are subject to similar interference effects.  As a result,
the effective $D^0$ branching fractions in this $D^0\bar D^0$ system differ
from those measured in isolated $D^0$ mesons.  Moreover, using time-independent
rate measurements, it becomes possible to probe $D^0$-$\bar D^0$ mixing as
well as the relative strong phases between $D^0$ and $\bar D^0$ decay
amplitudes to any given final state.

We implement the method described in
Ref.~\cite{Asner:2005wf} for measuring $y$ and $\delta$ using quantum
correlations at the $\psi(3770)$ resonance.
Our experimental technique is an extension of the double tagging method
previously used to determine absolute hadronic $D$-meson branching fractions
at CLEO\nobreakdash-c~\cite{dhad56}.
This method combines yields of fully reconstructed single tags (ST),
which are individually reconstructed $D^0$ or $\bar D^0$ candidates,
with yields of double tags (DT), which are events where both $D^0$ and
$\bar D^0$ are reconstructed, to
give absolute branching fractions without needing to know the luminosity or
$D^0\bar D^0$ production cross section.  Given a set of measured yields,
efficiencies, and background estimates, a least-squares
fitter~\cite{Sun:2005ip}
extracts the number of $D^0\bar D^0$ pairs produced (${\cal N}$) and the
branching fractions (${\cal B}$) of the reconstructed $D^0$ final states,
while accounting for all statistical and systematic uncertainties and their
correlations.
We employ a modified version of this fitter that also determines
$y$, $x^2$, $r^2$, $r\cos\delta$, and $rx\sin\delta$
using the following categories of
reconstructed final states: $K^-\pi^+$ ($f$),  $K^+\pi^-$ ($\bar f$),
$CP$-even ($S_+$) and $CP$-odd ($S_-$) eigenstates, and
semileptonic decays ($e^\pm$).  With CLEO\nobreakdash-c measurements alone,
$r^2$ is not determined with sufficient precision to extract $\delta$.
Therefore, we also
incorporate measurements of branching fractions and mixing parameters
from other CLEO\nobreakdash-c analyses or from external sources.
We neglect $CP$ violation in $D$ decays, which
would entail a slight correction to the mixing signal.

The paper is organized as follows.  In Section~\ref{sec:formalism}, we
review the formalism of quantum-correlated $D^0\bar D^0$ decay.
Section~\ref{sec:selection} describes the event selection criteria and
$D$ reconstruction procedures.  The external measurements used
in the fit are summarized in Section~\ref{sec:externalMeas}.
Systematic uncertainties, which are also input to the fit, are discussed
in Section~\ref{sec:systematics}.  Finally, we present and discuss our main
fit results in Section~\ref{sec:results}.  In the appendix, we provide
information for use by other experiments.


\section{Formalism}\label{sec:formalism}

To first order in $x$ and $y$, the $C$-odd rate $\Gamma_{D^0\bar D^0}(i,j)$
for $D^0\bar D^0$ decay to final state $\{i,j\}$ follows from the
anti-symmetric amplitude ${\cal M}_{ij}$:
\begin{eqnarray}
\nonumber
\Gamma_{D^0\bar D^0}(i,j) \propto {\cal M}^2_{ij} &=&
	\left|A_i \bar A_j - \bar A_i A_j \right|^2 \\
\label{eq:rates}
&=& \left|\langle i|D_2\rangle\langle j|D_1\rangle -
        \langle i|D_1\rangle\langle j|D_2\rangle \right|^2,
\end{eqnarray}
where $A_i\equiv\langle i|D^0\rangle$ and
$\bar A_i\equiv\langle i|\bar D^0\rangle$.
These amplitudes are normalized such that
${\cal B}_{K^-\pi^+}\approx A_{K^-\pi^+}^2(1+ry\cos\delta+rx\sin\delta)$,
${\cal B}_{S_\pm}\approx A_{S_\pm}^2(1\mp y)$, and
${\cal B}_e\approx A_e^2$.
The total rate, $\Gamma_{D^0\bar D^0}$, is the same as for uncorrelated decay,
as are ST rates.
However, unlike the case of uncorrelated $D^0\bar D^0$, we can consider the
$C$-odd $D^0\bar D^0$ system as a $D_1D_2$ pair.
If only flavored final states are considered, as in Ref.~\cite{dhad56},
then the effects of quantum correlations are negligible.  In this
analysis, we also include $CP$
eigenstates, which brings additional sensitivity to $y$ and $\delta$, as
demonstrated below.

Quantum-correlated semileptonic rates probe $y$ because the decay
width does not depend on the $CP$ eigenvalue of the parent $D$ meson,
as this weak decay is only sensitive to flavor content.  However, the total
width of the parent meson does depend on its $CP$ eigenvalue:
$\Gamma_{\shortstack[c]{{\scriptsize 1} \\ {\scriptsize 2}}}=\Gamma(1 \mp y)$,
so the semileptonic branching fraction
for $D_1$ or $D_2$ is modified by $1\pm y$.  If we reconstruct a semileptonic
decay in the
same event as a $D_2\to S_+$ decay, then the semileptonic $D$ must be a $D_1$.
Therefore, the effective quantum-correlated $D^0\bar D^0$ branching fractions
(${\cal F}^{\rm cor}$) for $\{S_\pm, e\}$ final states depend on $y$:
\begin{equation}
{\cal F}^{\rm cor}_{S_\pm, e} \approx
	2{\cal B}_{S_\pm}{\cal B}_{e}(1\pm y),
\end{equation}
where the factor of 2 arises from the sum of $e^+$ and $e^-$ rates.
Combined with estimates of ${\cal B}_e$ and ${\cal B}_{S_\pm}$
from ST yields, external sources, and flavor-tagged semileptonic yields,
this equation allows $y$ to be determined.

Similarly, if we reconstruct a $D\to K^-\pi^+$ decay in the same event
as a $D_2\to S_+$, then the $K^-\pi^+$ was produced from a $D_1$.
The effective branching fraction for this DT process is therefore
\begin{eqnarray}
\nonumber
{\cal F}^{\rm cor}_{S_+, K^-\pi^+} &=&
|\langle S_+|D_2\rangle \langle K^-\pi^+|D_1\rangle|^2 \\
\nonumber
&\approx& {\cal B}_{S_+}(1+y) |A_{K^-\pi^+} + \bar A_{K^-\pi^+}|^2 \\
\nonumber
&\approx& {\cal B}_{S_+}{\cal B}_{K^-\pi^+}(1+y)(1-ry\cos\delta-rx\sin\delta)|1+re^{-i\delta}|^2 \\
\nonumber
&=& {\cal B}_{S_+}{\cal B}_{K^-\pi^+}(1+y)(1-ry\cos\delta-rx\sin\delta)(1+2r\cos\delta+r^2)\\
&\approx& {\cal B}_{S_+}{\cal B}_{K^-\pi^+}( 1+2r\cos\delta+R_{\rm WS}+y ),
\end{eqnarray}
where $R_{\rm WS}$ is the wrong-sign rate ratio, which depends on $x$ and $y$
because of the interference between DCS and mixing transitions:
$R_{\rm WS} \equiv \Gamma(\bar D^0\to K^-\pi^+)/\Gamma(D^0\to K^-\pi^+) = r^2 + ry' + R_{\rm M}$.
In an analogous fashion, we find
${\cal F}^{\rm cor}_{S_-, K^-\pi^+}\approx{\cal B}_{S_-}{\cal B}_{K^-\pi^+}(1+R_{\rm WS}-2r\cos\delta-y)$.
When combined with knowledge of ${\cal B}_{S_+}$, $y$, and
$r$, the asymmetry between these two DT yields gives $\cos\delta$.
In the absence of quantum correlations, the effective branching fractions above
would be ${\cal B}_{S_\pm}{\cal B}_{K^-\pi^+}(1+R_{\rm WS})$.

More concretely, we evaluate Eq.~(\ref{eq:rates}) with
the above definitions of $r$
and $\delta$ to produce the expressions in Table~\ref{tab:rates}.
In doing so, we use the fact that inclusive ST rates
are given by the incoherent branching fractions since each event contains
one $D^0$ and one $\bar D^0$.
Comparison of ${\cal F}^{\rm cor}$ with the uncorrelated effective
branching fractions, ${\cal F}^{\rm unc}$, also given in
Table~\ref{tab:rates}, allows us to extract $r^2$, $r\cos\delta$, $y$, $x^2$,
and $rx\sin\delta$.
Information on ${\cal B}_i$ is obtained from ST yields at the $\psi(3770)$
and from external measurements using incoherently produced $D^0$ mesons.
These two estimates of ${\cal B}_i$ are averaged by the fitter to obtain
${\cal F}^{\rm unc}$.

\begin{table}[htb]
\caption{Correlated ($C$-odd) and uncorrelated effective $D^0\bar D^0$
branching fractions, ${\cal F}^{\rm cor}$ and
${\cal F}^{\rm unc}$, to leading order in $x$,
$y$ and $R_{\rm WS}$, divided by ${\cal B}_i$
for ST modes $i$ (first section) and
${\cal B}_i{\cal B}_j$ for DT modes $\{i,j\}$ (second section).
Charge conjugate modes are implied.}
\label{tab:rates}
\begin{tabular}{ccc}
\hline\hline
Mode & Correlated & Uncorrelated \\
\hline
$K^-\pi^+$ &
        $1+R_{\rm WS}$ &
        $1+R_{\rm WS}$ \\
$S_+$ & $2$ & $2$ \\
$S_-$ & $2$ & $2$ \\
\hline
$K^-\pi^+, K^-\pi^+$ &
        $R_{\rm M}$ &
        $R_{\rm WS}$ \\
$K^-\pi^+, K^+\pi^-$ &
	$(1+R_{\rm WS})^2-4r\cos\delta(r\cos\delta+y)$ &
        $1+R_{\rm WS}^2$ \\
$K^-\pi^+, S_+$ &
        $1+ R_{\rm WS}+2r\cos\delta+y$ &
        $1+R_{\rm WS}$ \\
$K^-\pi^+, S_-$ &
        $1+R_{\rm WS}- 2r\cos\delta-y$ &
        $1+R_{\rm WS}$ \\
$K^-\pi^+, e^-$ &
        $1-ry\cos\delta-rx\sin\delta$ &
        $1$ \\
$S_+, S_+$ & 0 & $1$ \\
$S_-, S_-$ & 0 & $1$ \\
$S_+, S_-$ &
        $4$ &
        $2$ \\
$S_+, e^-$ &
        $1+y$ &
        $1$ \\
$S_-, e^-$ &
        $1-y$ &
        $1$\\
\hline\hline
\end{tabular}
\end{table}

Using only ST and DT yields at the $\psi(3770)$, we can determine $y$ and
$\cos\delta$ from the following double ratios, obtained by manipulating
the expressions in Table~\ref{tab:rates}:
\begin{eqnarray}
\label{eq:y1}
y &\approx& \frac{N_{fe}}{4N_f}\left[
       \frac{N_{S_-}}{N_{S_-e}} - \frac{N_{S_+}}{N_{S_+e}} \right]\\
\label{eq:y1a}
  &\approx& \frac{1}{4}\left[
	\frac{N_{S_+e}N_{S_-}}{N_{S_-e}N_{S_+}} -
	\frac{N_{S_-e}N_{S_+}}{N_{S_+e}N_{S_-}} \right] \\
\label{eq:rz1}
2r\cos\delta+y &\approx& \frac{N_{f\bar f}}{4N_{\bar f}}\left[
	\frac{N_{S_-}}{N_{S_-f}} - \frac{N_{S_+}}{N_{S_+f}}\right]\\
\label{eq:rz1a}
&\approx& \frac{1}{4}N_f N_{S_+S_-}\left[
	\frac{1}{N_{fS_-}N_{S_+}} - \frac{1}{N_{fS_+}N_{S_-}} \right],
\end{eqnarray}
where $N$ denotes an efficiency-corrected background-subtracted ST or DT yield.
Note that
semileptonic yields are essential for separating $y$ and $\cos\delta$.
Including external branching fractions provides
additional ways to determine $y$ and $\cos\delta$:
\begin{eqnarray}
\label{eq:y2}
y &=& \pm \left[1 -
	\frac{2{\cal N}{\cal B}_{S_\pm}{\cal B}_{e}}{N_{S_\pm e}}\right]\\
\label{eq:rz2}
2r\cos\delta+y &=& (1+R_{\rm WS})\left[
       \frac{N_{S_+f}/{\cal B}_{S_+}-N_{S_-f}/{\cal B}_{S_-}}
       {N_{S_+f}/{\cal B}_{S_+}+{N_{S_-f}/\cal B}_{S_-}}\right].
\end{eqnarray}
Although we neglect $x^2$ and $y^2$ terms in general, we report a result for
$x^2$ as determined solely from $N_{ff}/N_{f\bar f}$.


\section{Event Selection and Reconstruction}~\label{sec:selection}

Our current analysis uses 281 ${\rm pb}^{-1}$ of $e^+e^-\to\psi(3770)$ data
collected with the CLEO\nobreakdash-c~\cite{cleo2,cleo2.5,cleo3,rich,cleo-c}
detector.
We also make use of a large Monte Carlo simulated sample of uncorrelated
$D^0\bar D^0$ decays with an effective luminosity 40 times that of our data
sample, from which we estimate signal efficiencies, background contributions,
and probabilities for misreconstructing a produced signal decay in a
different signal mode (crossfeed).
In these samples, we reconstruct the final states shown in
Table~\ref{tab:finalStates},
with $\pi^0\to\gamma\gamma$, $\eta\to\gamma\gamma$, $K^0_S\to\pi^+\pi^-$, and
$\omega\to\pi^+\pi^-\pi^0$.
Because most $K^0_L$ mesons and neutrinos are not detected, we do not
reconstruct $K^0_L\pi^0$ and semileptonic ST modes; they are only included
in DT modes, paired with a fully reconstructed $D$ candidate.
Below, we denote by $S_+'$ the subset of $S_+$ modes that are
fully reconstructed: $K^+K^-$, $\pi^+\pi^-$, and $K^0_S\pi^0\pi^0$.
For $CP$ eigenstates, we choose modes with unambiguous $CP$ content.
In addition to two-body decays, we also include
$K^0_S\pi^0\pi^0$, which is a pure $CP$-even eigenstate because the
two identical $\pi^0$'s must have even angular momentum in order to satisfy
Bose symmetry.  We neglect $CP$ violation in $K^0$ decays.

\begin{table}[htb]
\caption{$D$ final states reconstructed in this analysis.}
\label{tab:finalStates}
\begin{tabular}{cc}
\hline\hline
Type & Final States \\
\hline
Flavored & $K^-\pi^+$, $K^+\pi^-$ \\
$S_+$ & $K^+K^-$, $\pi^+\pi^-$, $K^0_S\pi^0\pi^0$, $K^0_L\pi^0$ \\
$S_-$ & $K^0_S\pi^0$, $K^0_S\eta$, $K^0_S\omega$ \\
$e^\pm$ & Inclusive $Xe^+\nu_e$, $Xe^-\bar\nu_e$ \\
\hline\hline
\end{tabular}
\end{table}

Standard CLEO\nobreakdash-c selection critera for
$\pi^\pm$, $K^\pm$, $\pi^0$, and
$K^0_S$ candidates are described in Ref.~\cite{dhad281}.
In addition, for $K^0_S$ candidates, we impose
$|M(\pi^+\pi^-)-M_{K^0_S}|<7.5$ MeV/$c^2$, and we require the decay vertex to
be separated from the interaction region with a significance greater than
two standard deviations ($\sigma$).  For the $K^0_S\pi^0\pi^0$ mode, we
reject $K^0_S$ daughter tracks with ionization energy loss consistent with
being kaons at the level of $2\sigma$ in order to suppress misreconstructed
$D\to K^-\pi^+\pi^0$ decays, where the kaon is taken to be a charged pion,
and a soft combinatoric $\pi^0$ candidate is incorporated into the
$D$ candidate.
We accept $\omega$ candidates with
$|M(\pi^+\pi^-\pi^0)-M_\omega| < 20$ MeV/$c^2$.  Reconstruction
of $\eta\to\gamma\gamma$ proceeds analogously to $\pi^0\to\gamma\gamma$.
In addition, we require $|M(\gamma\gamma)-M_\eta| < 42$ MeV/$c^2$.
All invariant mass requirements correspond to approximately $3\sigma$
consistency with the nominal masses from Ref.~\cite{pdg06}.

\subsubsection{Single Tags}\label{sec:st}

Reconstruction of all modes in this analysis, including DT modes, begins with
fully reconstructed ST $D$ candidates, which are identified using two kinematic
variables that express momentum and energy conservation: the beam-constrained
candidate mass $M$ and the energy difference $\Delta E$.  These
variables are defined to be
\begin{eqnarray}
M &\equiv& \sqrt{ E_0^2 / c^4 - {\bf p}_D^2 / c^2 } \\
\Delta E &\equiv& E_D - E_0,
\end{eqnarray}
where ${\bf p}_D$ and $E_D$ are the total momentum and energy of the $D$
candidate,
and $E_0$ is the beam energy.  Correctly reconstructed $D$ candidates produce
a peak in $M$ at the $D$ mass and in $\Delta E$ at zero.
We determine ST yields by fitting the $M$ distribution with the
mode-dependent requirements on $\Delta E$ listed in Table~\ref{tab:DeltaEcuts},
which are applied to both ST and DT $D$ candidates.  The limits are set at
approximately three standard deviations.  Modes with $\pi^0$ or $\eta$, which
decay to two photons, have asymmetric limits to allow for partially contained
showers in the electromagnetic calorimeter.

\begin{table}[htb]
\caption{Requirements on $\Delta E$ for $D$ candidates.
\label{tab:DeltaEcuts}}
\begin{tabular}{lc}\hline\hline
Mode              &  Requirement (GeV)\\ \hline
$K^-\pi^+$        &  $|\Delta E|<0.0294$      \\
$K^+\pi^-$        &  $|\Delta E|<0.0294$      \\
$K^+K^-$          &  $|\Delta E|<0.0200$    \\
$\pi^+\pi^-$      &  $|\Delta E|<0.0300$    \\
$K^0_S\pi^0\pi^0$ &  $-0.0550<\Delta E<0.0450$    \\
$K^0_S\pi^0$      &  $-0.0710<\Delta E<0.0450$    \\
$K^0_S\eta$       &  $-0.0550<\Delta E<0.0350$    \\
$K^0_S\omega$     &  $|\Delta E|<0.0250$    \\
\hline\hline
\end{tabular}
\end{table}

For $K^\pm\pi^\mp$, $K^+K^-$, and $\pi^+\pi^-$ ST modes, in events
containing only two tracks, we suppress cosmic muons and Bhabhas scattering
events by vetoing tracks that are identified as muons or electrons and
by requiring at least one electromagnetic shower in the calorimeter above
50 MeV not associated with the signal tracks.
For $K^+K^-$ ST candidates, additional geometric requirements are needed to
remove
doubly radiative Bhabhas followed by pair conversion of a radiated photon.
Also, we accept only one candidate per mode per event;
when multiple candidates are present, we choose the one with smallest
$|\Delta E|$.

\begin{figure*}[htb]
\includegraphics*[width=\linewidth]{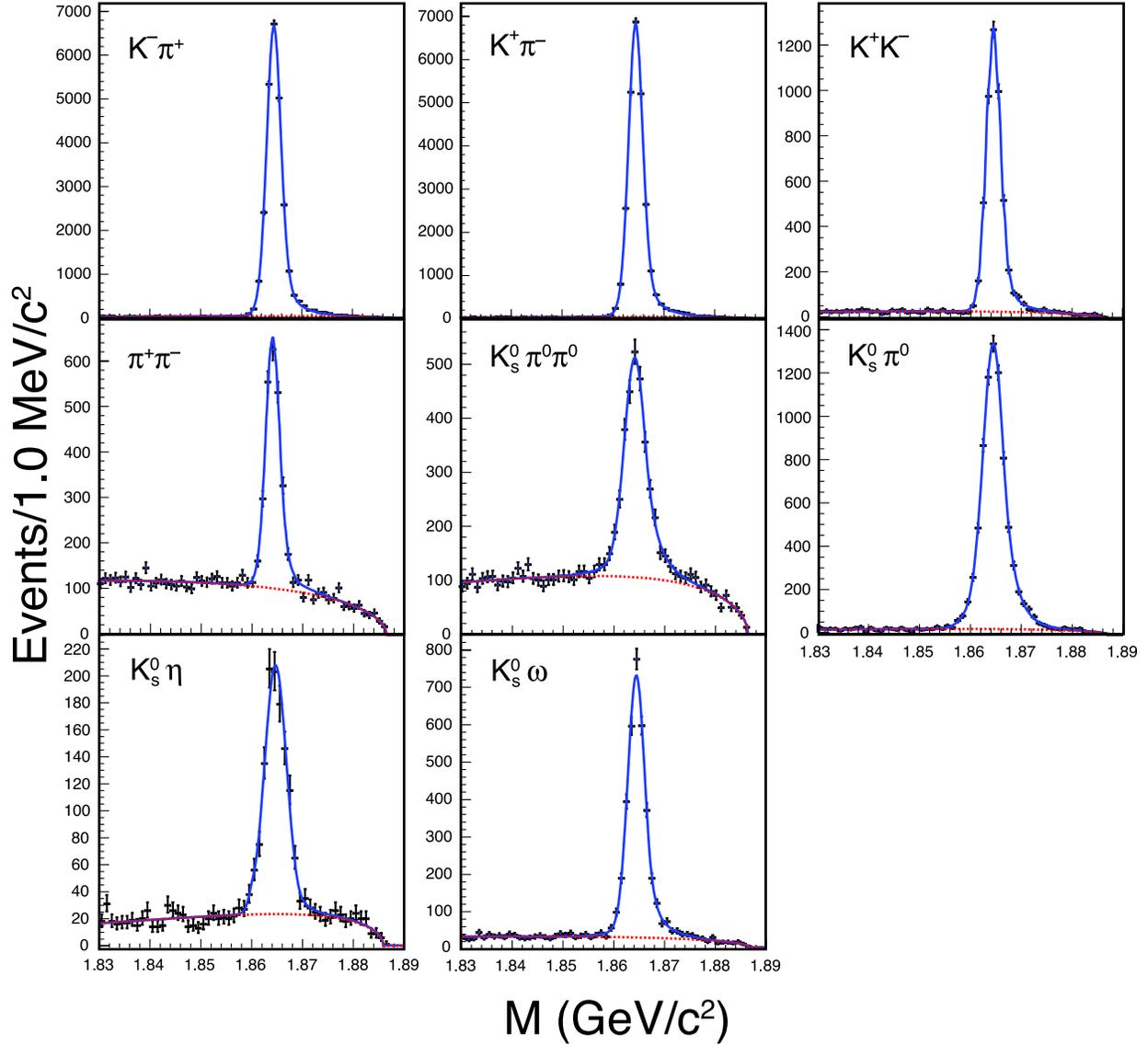}
\caption{
ST $M$ distributions and fits.
Data are shown as points with error bars.  The solid lines show
the total fits, and the dashed lines show the background shapes.
}
\label{fig:st}
\end{figure*}

The resultant $M$ distributions, shown in Fig.~\ref{fig:st},
are fitted to a signal shape derived from simulated signal events and to a
background ARGUS function~\cite{argus}.
The simulated signal shape is both shifted and convoluted
with a Gaussian smearing function to account for imperfect modeling of the
detector resolution and beam energy calibration.  The width of the smearing
function is allowed to float in the fit.  The measured ST yields and
efficiencies are given in Table~\ref{tab:STYieldEffs}.  The yield
uncertainties are statistical and uncorrelated systematic, respectively.
The latter arise from modeling of multiple candidates in simulation and
variations in the signal lineshape.  Correlated systematic uncertainties
are discussed separately in Section~\ref{sec:systematics}.

\begin{table}[htbp]
\begin{center}
\caption{ST yields and efficiencies input to the data fit.  Yield
uncertainties are statistical and uncorrelated systematic, respectively,
and efficiency uncertainties are statistical only.}
\label{tab:STYieldEffs}
\begin{tabular}{lcc}
\hline\hline
Mode &  ~~Yield~~ & ~~Efficiency (\%) \\ \hline
$K^-\pi^+$ &
        $25374\pm 166\pm 26$ &
        $64.70\pm 0.04$ \\
$K^+\pi^-$ &
        $25842\pm 167\pm 26$ &
        $65.62\pm 0.04$ \\
$K^+K^-$ &
        $4740\pm 71\pm 5$ &
        $57.25\pm 0.09$ \\
$\pi^+\pi^-$ &
        $2098\pm 59\pm 9$ &
        $72.92\pm 0.13$ \\
$K^0_S\pi^0\pi^0$ &
        $2435\pm 72\pm 16$ &
        $12.50\pm 0.06$ \\
$K^0_S\pi^0$ &
        $7523\pm 91\pm 17$ &
        $29.73\pm 0.05$ \\
$K^0_S\eta$ &
        $1051\pm 39\pm 17$ &
        $10.34\pm 0.06$ \\
$K^0_S\omega$ &
        $3239\pm 63\pm 7$ &
        $12.48\pm 0.04$ \\
\hline\hline
\end{tabular}
\end{center}
\end{table}

\subsubsection{Fully Reconstructed Hadronic Double Tags}

\begin{figure}[htb]
\includegraphics*[width=\linewidth]{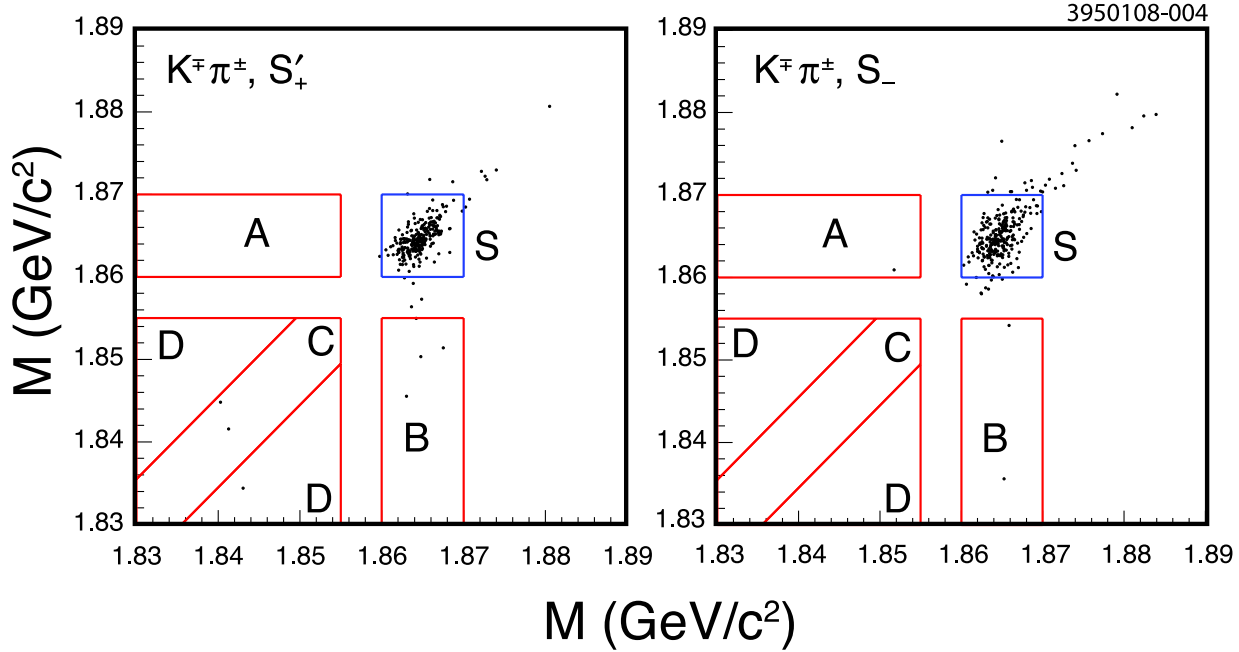}
\caption{Two-dimensional $M$ distributions with signal (S) and
sideband (A, B, C, D) regions depicted, for the sum of all
$\{K^\mp\pi^\pm, S_+'\}$ and $\{K^\mp\pi^\pm, S_-\}$ modes listed in
Table~\ref{tab:DTYieldEffs}.
}
\label{fig:dtRegions}
\end{figure}

Except for modes with $K^0_L\pi^0$, we form hadronic DTs by combining
two ST candidates passing
the above selection criteria.  Multiple candidates are resolved after
forming the DT candidates, not at the ST level.  We choose
one candidate per mode per event with $\bar M$ closest to the measured $D^0$
mass, where $\bar M\equiv [M(D^0) + M(\bar D^0)]/2$.
We extract signal yields by counting events in the
two-dimensional $M(D^0)$ vs. $M(\bar D^0)$ plane, as
illustrated in Fig.~\ref{fig:dtRegions}.
The signal region S is defined to be approximately three standard deviations
in each dimension:
$1.86 < M(D^0) < 1.87$ GeV/$c^2$ and $1.86 < M(\bar D^0)<1.87$ GeV/$c^2$.
Sidebands A and B contain candidates where either the $D^0$ or the $\bar D^0$
is misreconstructed.  Sidebands C and D contain candidates where both
$D^0$ and $\bar D^0$ are misreconstructed, either in a correlated way (C)
by assigning daughter particles to the wrong parent or in an uncorrelated
way (D).  Event counts in sidebands A, B, and C are projected into the signal
region S using scale factors determined from integrating the
background shape in the ST $M$ fits.  Contributions to sideband D are
assumed to be uniformly distributed across the other regions.  To account for
systematic effects in the sideband definitions and in the extrapolation to the
signal regions, we assign a 100\% systematic uncertainty on the size of the
sideband subtractions, which is much smaller than the statistical
uncertainties in all cases.

\begin{figure*}[htb]
\includegraphics*[width=\linewidth]{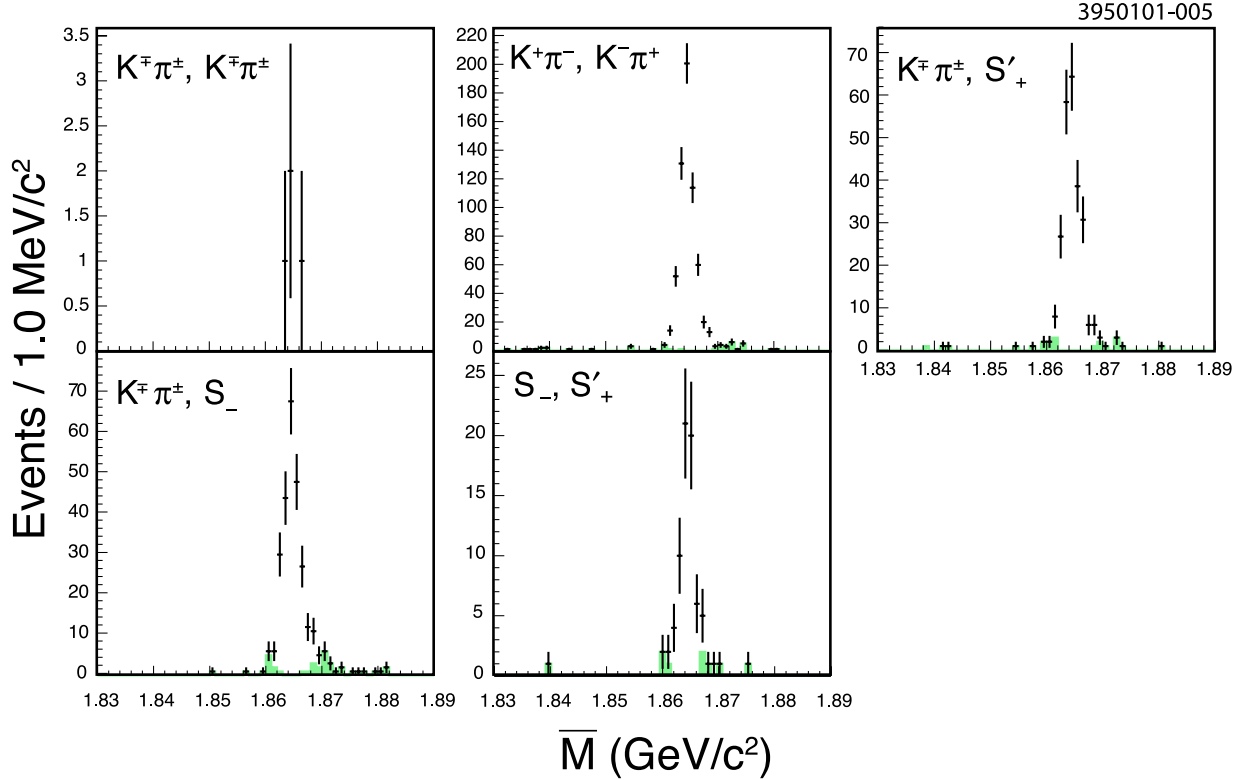}
\caption{
Sums of fully reconstructed DT $\bar M$ distributions, with charge conjugate
modes implied.
Data are shown as points with error bars.  The shaded histograms show events
outside the signal region.
}
\label{fig:dtHad}
\end{figure*}

Table~\ref{tab:DTYieldEffs} gives the fully reconstructed DT yields and
efficiencies input to the fit, and Fig.~\ref{fig:dtHad} shows the
corresponding $\bar M$ projections.  Same-$CP$ $\{S_\pm,S_\pm\}$ modes are not
included in the standard fit.

\begin{table}[htbp]
\begin{center}
\caption{Fully reconstructed DT yields and efficiencies input to the data fit.
Yield uncertainties are statistical and uncorrelated systematic
(for sideband subtraction), respectively,
and efficiency uncertainties are statistical only.}
\label{tab:DTYieldEffs}
\begin{tabular}{lcc}
\hline\hline
Mode & Yield & Efficiency (\%) \\
\hline
$K^-\pi^+, K^-\pi^+$ &
        $2.0\pm 1.4\pm 0$ &
        $36.1\pm 3.4$ \\
$K^-\pi^+, K^+\pi^-$ &
        $600\pm 25\pm 5$ &
        $41.1\pm 0.2$ \\
$K^-\pi^+, K^+K^-$ &
        $71\pm 8\pm 1$ &
        $35.5\pm 0.6$ \\
$K^-\pi^+, \pi^+\pi^-$ &
        $24\pm 5\pm 1$ &
        $44.4\pm 1.1$ \\
$K^-\pi^+, K^0_S\pi^0\pi^0$ &
        $32\pm 6\pm 1$ &
        $8.0\pm 0.3$ \\
$K^-\pi^+, K^0_S\pi^0$ &
        $88\pm 9\pm 1$ &
        $18.4\pm 0.3$ \\
$K^-\pi^+, K^0_S\eta$ &
        $8.0\pm 2.8\pm 0.0$ &
        $6.0\pm 0.3$ \\
$K^-\pi^+, K^0_S\omega$ &
        $29\pm 5\pm 0$ &
        $8.7\pm 0.2$ \\
$K^+\pi^-, K^+\pi^-$ &
        $2.0\pm 1.4\pm 0.0$ &
        $44.1\pm 3.3$ \\
$K^+\pi^-, K^+K^-$ &
        $54\pm 7\pm 0$ &
        $36.1\pm 0.6$ \\
$K^+\pi^-, \pi^+\pi^-$ &
        $25\pm 5\pm 1$ &
        $48.1\pm 1.1$ \\
$K^+\pi^-, K^0_S\pi^0\pi^0$ &
        $33\pm 6\pm 0$ &
        $8.0\pm 0.3$ \\
$K^+\pi^-, K^0_S\pi^0$ &
        $76\pm 9\pm 0$ &
        $18.6\pm 0.3$ \\
$K^+\pi^-, K^0_S\eta$ &
        $9\pm 3\pm 0$ &
        $6.1\pm 0.3$ \\
$K^+\pi^-, K^0_S\omega$ &
        $33\pm 6\pm 1$ &
        $8.0\pm 0.2$ \\
$K^+K^-, K^0_S\pi^0$ &
        $39\pm 6\pm 1$ &
        $17.1\pm 0.6$ \\
$K^+K^-, K^0_S\eta$ &
        $7.0\pm 2.7\pm 0.0$ &
        $7.1\pm 0.8$ \\
$K^+K^-, K^0_S\omega$ &
        $20\pm 4\pm 0$ &
        $6.8\pm 0.4$ \\
$\pi^+\pi^-, K^0_S\pi^0$ &
        $13\pm 4\pm 0$ &
        $19.2\pm 1.2$ \\
$\pi^+\pi^-, K^0_S\eta$ &
        $2.0\pm 1.4\pm 0.0$ &
        $8.1\pm 1.4$ \\
$\pi^+\pi^-, K^0_S\omega$ &
        $7.0\pm 2.7\pm 0.0$ &
        $9.9\pm 0.9$ \\
$K^0_S\pi^0\pi^0, K^0_S\pi^0$ &
        $14\pm 4\pm 0$ &
        $3.5\pm 0.2$ \\
$K^0_S\pi^0\pi^0, K^0_S\eta$ &
        $4.0\pm 2.0\pm 0.0$ &
        $1.2\pm 0.2$ \\
$K^0_S\pi^0\pi^0, K^0_S\omega$ &
        $4.0\pm 2.0\pm 0.0$ &
        $1.4\pm 0.2$ \\
\hline\hline
\end{tabular}
\end{center}
\end{table}

\subsubsection{Semileptonic Double Tags}

Semileptonic DTs are partially reconstructed by combining a
fully reconstructed hadronic ST with an electron candidate from the
remainder of the event.  The hadronic
tags are selected using the same criteria as for STs, described in
in Section~\ref{sec:st}, with an additional mode-dependent $M$
requirement listed in Table~\ref{tab:slMBCCuts}.  The limits are set at
approximately three standard deviations.
Multiple tag candidates are resolved with minimal $|\Delta E|$, as for STs.
All electron candidates in a given event are accepted.

\begin{table}[htb]
\caption{Requirements on $M$ for hadronic tags in semileptonic
DT candidates
\label{tab:slMBCCuts}}
\begin{tabular}{lc}\hline\hline
Mode              &  Requirement (GeV/$c^2$)\\ \hline
$K^-\pi^+$        &  $1.8585 < M < 1.8775$      \\
$K^+\pi^-$        &  $1.8585 < M < 1.8775$    \\
$K^+K^-$          &  $1.8585 < M < 1.8775$   \\
$\pi^+\pi^-$      &  $1.8585 < M < 1.8775$   \\
$K^0_S\pi^0\pi^0$ &  $1.8530 < M < 1.8780$   \\
$K^0_S\pi^0$      &  $1.8530 < M < 1.8780$   \\
$K^0_S\eta$       &  $1.8585 < M < 1.8775$    \\
$K^0_S\omega$     &  $1.8585 < M < 1.8775$  \\
\hline\hline
\end{tabular}
\end{table}

Electron candidate tracks are selected using the criteria described in
Ref.~\cite{eid}.  Electrons are distinguished from hadrons via a multivariate
discriminant that combines information from the ratio of the
energy deposited in the calorimeter to the measured
track momentum ($E/p$), ionization energy loss in the tracking chambers
($dE/dx$), and the ring-imaging \v{C}erenkov counter (RICH).
When paired with a $K^\mp\pi^\pm$ tag, the kaon and electron charges must be
the same.  This requirement removes charge-symmetric backgrounds and
cannot be used with unflavored $S_\pm$ tag modes.

\begin{figure}[htb]
\includegraphics*[width=\linewidth]{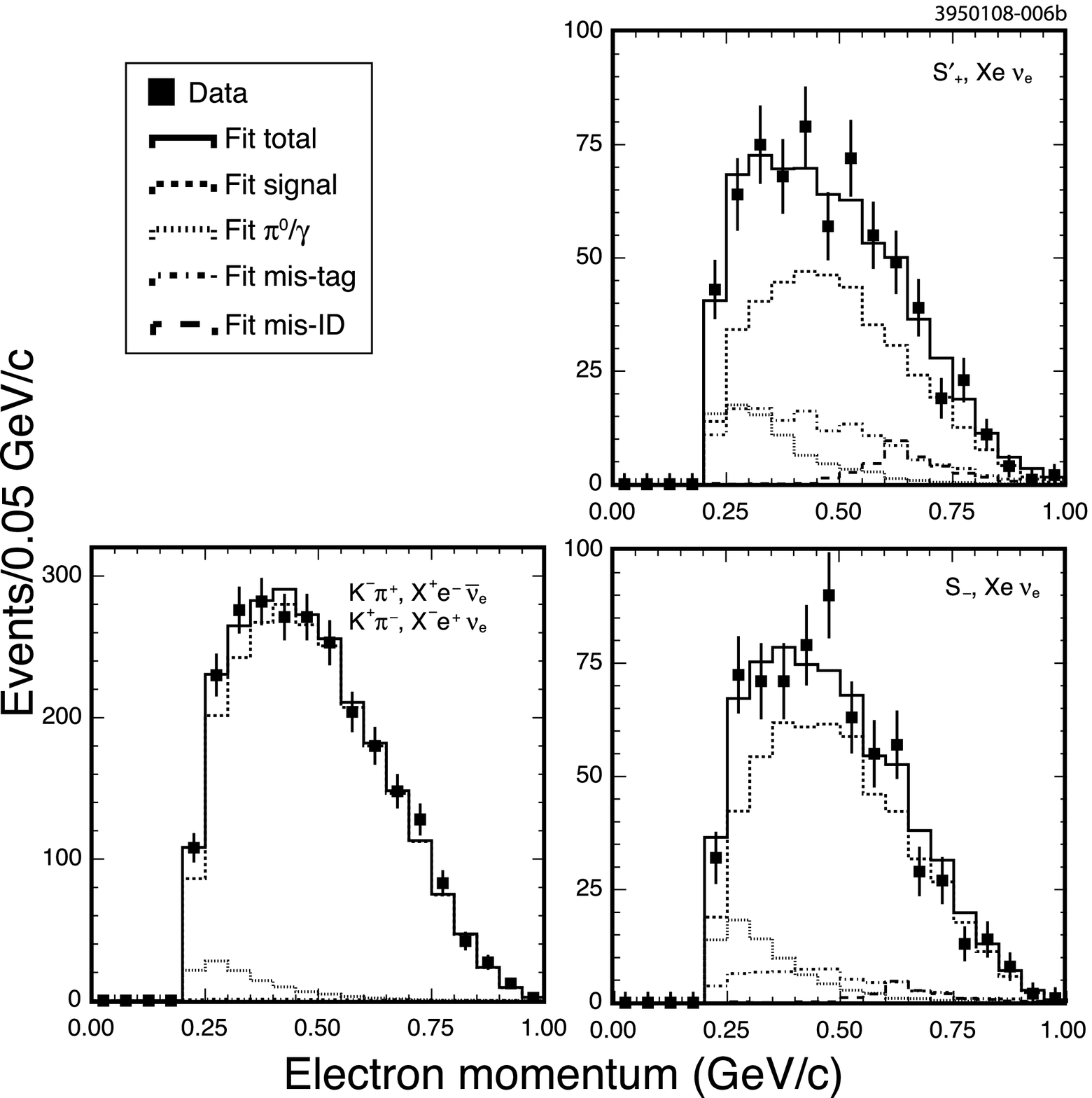}
\caption{Sums of electron momentum spectra for $X^-e^+\nu_e$ and
$X^+e^-\bar\nu_e$.  Data are shown as points with error bars, and
the histograms show the signal and background components in the fits.}
\label{fig:dtSL}
\end{figure}

As illustrated in Fig.~\ref{fig:dtSL}, the resulting electron momentum
spectrum for each tag mode 
is fit to signal and background shapes fixed by the simulation; only the
normalizations are allowed to float.
We correct the simulated signal spectrum for relative bin-by-bin efficiency
differences
between data and simulation, which are measured with a high-statistics
radiative Bhabha sample.
The background components in the spectrum fit include misreconstructed tags,
photon conversions
and light hadron decays to electrons (mostly $\pi^0$ Dalitz decays),
weak decays in flight, and hadron misidentification.  The latter component
peaks at 700 MeV/$c$ and mostly consists of $K^\pm$ that escape the
acceptance of the RICH; $K^\pm$ are
indistinguishable from $e^\pm$ using $dE/dx$ at this momentum.
In the fit, the misreconstructed tag component is fixed from sidebands in
$M$ and $\Delta E$.  The other backgrounds contribute
approximately 7\% for $K^\mp\pi^\pm$ tags and 20--30\% for $S_\pm$ tags.

Table~\ref{tab:SLYieldEffs} gives the semileptonic DT yields and efficiencies
input to the fit.  The uncorrelated systematic uncertainties are determined
from yield excursions under variation of the signal and background shapes
used in the spectrum fit.  For the signal
shape, we adjust the semileptonic form factor model used in the simulation.
For the background shapes, we vary the allowed background composition by
removing components with insignificant yields and also allowing for negative
normalizations.

\begin{table}[htbp]
\begin{center}
\caption{Semileptonic DT yields and efficiencies input to the data fit.
Yield uncertainties are statistical and uncorrelated systematic, respectively,
and efficiency uncertainties are statistical only.}
\label{tab:SLYieldEffs}
\begin{tabular}{lcc}
\hline\hline
Mode & Yield & Efficiency (\%) \\
\hline
$X^+e^-\bar\nu_e, K^-\pi^+$ &
        $1128\pm 45\pm 13$ &
        $45.3\pm 0.2$ \\
$X^+e^-\bar\nu_e, K^-K^+$ &
        $128\pm 24\pm 5$ &
        $40.1\pm 0.5$ \\
$X^+e^-\bar\nu_e, \pi^-\pi^+$ &
        $49\pm 12\pm 7$ &
        $50.3\pm 0.9$ \\
$X^+e^-\bar\nu_e, K^0_S\pi^0\pi^0$ &
        $37\pm 22\pm 12$ &
        $8.9\pm 0.2$ \\
$X^+e^-\bar\nu_e, K^0_S\pi^0$ &
        $195\pm 24\pm 2$ &
        $21.4\pm 0.3$ \\
$X^+e^-\bar\nu_e, K^0_S\eta$ &
        $28\pm 6\pm 4$ &
        $7.9\pm 0.3$ \\
$X^+e^-\bar\nu_e, K^0_S\omega$ &
        $50\pm 15\pm 6$ &
        $8.2\pm 0.2$ \\
$X^-e^+\nu_e, K^+\pi^-$ &
        $1218\pm 47\pm 17$ &
        $45.9\pm 0.2$ \\
$X^-e^+\nu_e, K^-K^+$ &
        $102\pm 21\pm 10$ &
        $39.0\pm 0.5$ \\
$X^-e^+\nu_e, \pi^-\pi^+$ &
        $40\pm 10\pm 4$ &
        $50.5\pm 0.8$ \\
$X^-e^+\nu_e, K^0_S\pi^0\pi^0$ &
        $50\pm 15\pm 9$ &
        $9.6\pm 0.2$ \\
$X^-e^+\nu_e, K^0_S\pi^0$ &
        $189\pm 19\pm 7$ &
        $21.4\pm 0.3$ \\
$X^-e^+\nu_e, K^0_S\eta$ &
        $27\pm 8\pm 2$ &
        $7.7\pm 0.3$ \\
$X^-e^+\nu_e, K^0_S\omega$ &
        $49\pm 16\pm 4$ &
        $8.5\pm 0.2$ \\
\hline\hline
\end{tabular}
\end{center}
\end{table}

\subsubsection{Double Tags with $K^0_L\pi^0$}

DT modes with $K^0_L\pi^0$ are reconstructed with the missing mass technique
also used in Ref.~\cite{ksklpi}.
A fully reconstructed ST candidate, selected as described in
Section~\ref{sec:st}, is combined with a $\pi^0$ candidate, and we compute
the squared recoil mass against the ST-$\pi^0$ system, $M^2_{\rm miss}$.
Signal $K^0_L\pi^0$ decays produce a peak in $M^2_{\rm miss}$ at $M_{K^0_L}$.
Backgrounds from $D\to K^0_S\pi^0$, $\pi^0\pi^0$, and $\eta\pi^0$ are
suppressed by vetoing events with additional unassigned charged particles,
$\eta$, or $\pi^0$ candidates.

\begin{figure}[htb]
\includegraphics*[width=\linewidth]{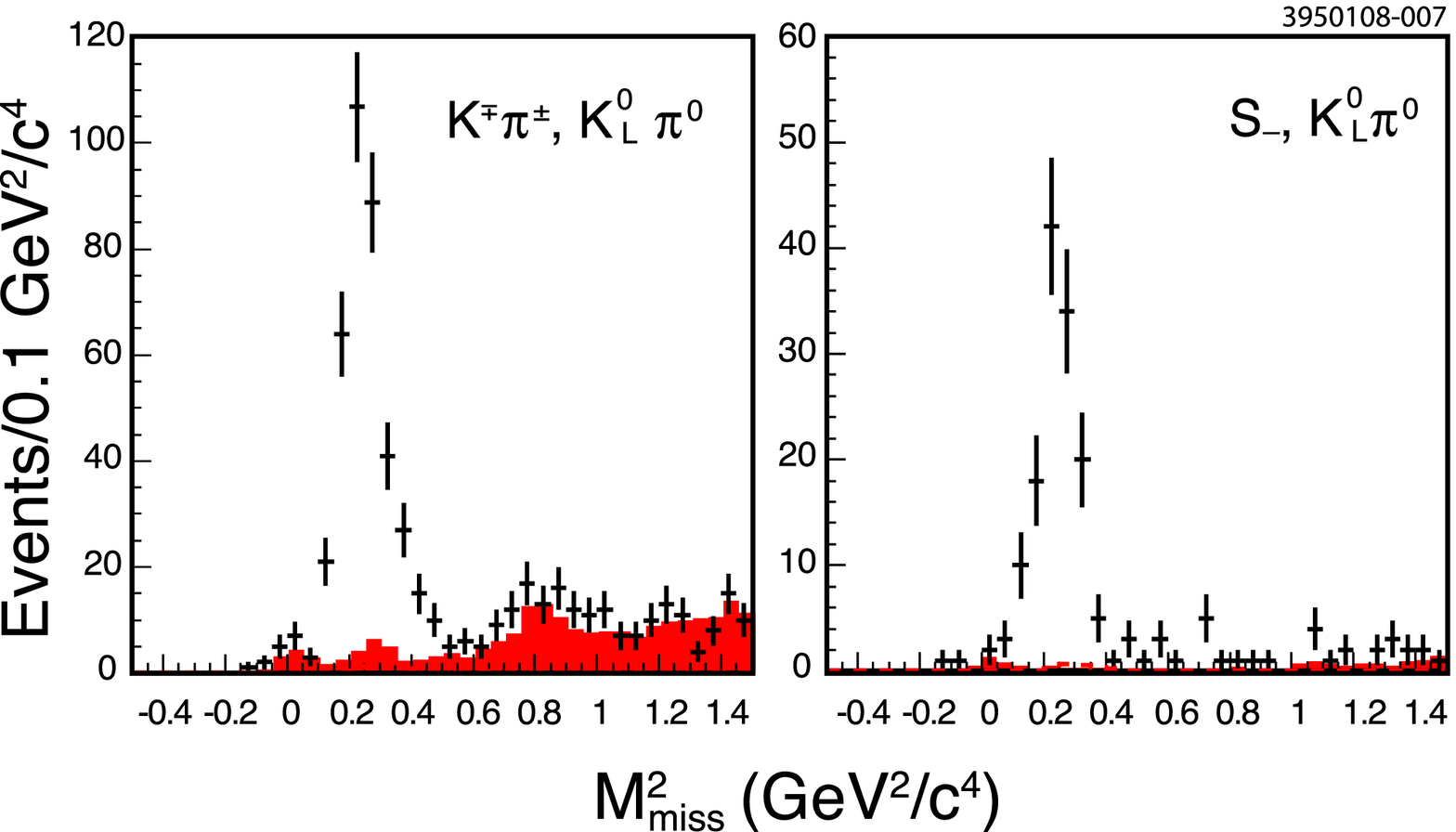}
\caption{
Sums of $M^2_{\rm miss}$ distributions for $K^0_L\pi^0$.
The shaded histograms represent simulations of the peaking
backgrounds $D^0\to\pi^0\pi^0$, $K^0_S\pi^0$,
$\eta\pi^0$, and $K^{*0}\pi^0$.
}
\label{fig:dtK0L}
\end{figure}

Figure~\ref{fig:dtK0L} shows the resultant $M^2_{\rm miss}$ in data.
Signal yields are obtained from event counts in the signal region of
$0.1 < M^2_{\rm miss} < 0.5$ ${\rm GeV}^2/c^4$.  The contribution from
combinatoric background is estimated from the
$0.8 < M^2_{\rm miss} < 1.2$ ${\rm GeV}^2/c^4$ sideband to be
${\cal O}(10^{-3})$
and is subtracted.  The residual background contributions from
$K^0_S\pi^0$ and $\eta\pi^0$, which peak in the signal region, and from
$\pi^0\pi^0$, which peaks below the signal region, are estimated from 
simulated data and are subtracted by the fitter, as described in
Section~\ref{sec:backgrounds}.

Table~\ref{tab:K0LYieldEffs} lists the $K^0_L\pi^0$ DT yields and efficiencies
input to the fit.  There are no uncorrelated systematic uncertainties for
these modes.

\begin{table}[htbp]
\begin{center}
\caption{$K^0_L\pi^0$ DT yields and efficiencies with statistical
uncertainties input to the data fit.}
\label{tab:K0LYieldEffs}
\begin{tabular}{lcc}
\hline\hline
Mode & Yield & Efficiency (\%) \\
\hline
$K^0_L\pi^0, K^-\pi^+$ &
        $187\pm 14$ &
        $34.0\pm 0.2$ \\
$K^0_L\pi^0, K^+\pi^-$ &
        $179\pm 14$ &
        $34.8\pm 0.2$ \\
$K^0_L\pi^0, K^0_S\pi^0$ &
        $90\pm 10$ &
        $15.5\pm 0.1$ \\
$K^0_L\pi^0, K^0_S\eta$ &
        $8.6\pm 3.1$ &
        $5.4\pm 0.1$ \\
$K^0_L\pi^0, K^0_S\omega$ &
        $34\pm 6$ &
        $6.2\pm 0.1$ \\
\hline\hline
\end{tabular}
\end{center}
\end{table}

\subsubsection{Crossfeed and Peaking Backgrounds}~\label{sec:backgrounds}

The yields in Tables~\ref{tab:STYieldEffs}, \ref{tab:DTYieldEffs},
\ref{tab:SLYieldEffs}, and~\ref{tab:K0LYieldEffs} include peaking backgrounds
contributions and crossfeed among signal modes that are subtracted by the
fitter.  To do so, the fitter makes use of
crossfeed probabilities and background efficiencies determined
from simulated events, as well as
branching fractions for peaking background processes~\cite{pdg06}.
These inputs are listed in Table~\ref{tab:backgroundSources}.
The correlated uncertainties among these contributions is also accounted for.

In most cases, the peaking background processes produce the same final state
particles as the signal modes to which they contribute.  For these backgrounds,
the DT contribution is assumed to occur at the same rate as for STs.
Other backgrounds, indicated by asterisks in Table~\ref{tab:backgroundSources},
arise from misreconstructed $D$ decays.  These backgrounds are present only
in ST modes and do not contribute to DT modes because of the kinematic
constraints of full event reconstruction.
Background processes are identified using simulated $D^0\bar D^0$ events, where
each $D$ is uncorrelated with the other and decays generically.  Background
efficiencies are determined by observing the change in signal yield as
each background contribution is removed from the simulated sample.

\begin{table}[htbp]
\begin{center}
\caption{Crossfeed probabilities and peaking background efficiencies and
branching fractions input to the fit. Backgrounds marked by
an asterisk (*) occur in only in STs, not DTs.}
\label{tab:backgroundSources}
\begin{tabular}{rclcc}
\hline\hline
Crossfeed &$\to$& Signal & & Probability (\%) \\
\hline
$K^+\pi^-$ &$\to$& $K^-\pi^+$ & & $0.098\pm 0.005$ \\
$K^-\pi^+$ &$\to$& $K^+\pi^-$ & & $0.092\pm 0.003$ \\
$K^0_S\omega$ &$\to$& $K^0_S\pi^0\pi^0$ & & $0.048\pm 0.031$ \\
$K^0_S\pi^0$ &$\to$& $K^0_L\pi^0$ & & $1.92\pm 0.03$ \\
\hline
Background &$\to$& Signal & ${\cal B}_{\rm bkg}$ (\%)~\cite{pdg06} &
        Efficiency (\%) \\
\hline
$K^0_S\pi^+\pi^-$ &$\to$& $K^0_S\pi^0\pi^0$ & 
        $2.90\pm 0.19$ & $0.006\pm 0.012$ \\
$\pi^+\pi^-\pi^0\pi^0$ &$\to$& $K^0_S\pi^0\pi^0$ & 
        $0.98\pm 0.09$ & $0.029\pm 0.067$ \\
$K^-\pi^+\pi^0$ (*) &$\to$& $K^0_S\pi^0\pi^0$ & 
        $14.1\pm 0.5$ & $0.0012\pm 0.0024$ \\
$D^+\to K^0_S\pi^+\pi^0$ (*) &$\to$& $K^0_S\pi^0\pi^0$ & 
        $7.22\pm 0.26$ & $0.048\pm 0.007$ \\
$\rho^+\pi^-$ &$\to$& $K^0_S\pi^0$ & 
        $1.00\pm 0.06$ & $0.071\pm 0.005$ \\
$\rho^0\pi^0$ &$\to$& $K^0_S\pi^0$ & 
        $0.32\pm 0.04$ & $0.032\pm 0.033$ \\
Generic $D^0\bar D^0$ (*) &$\to$& $K^0_S\pi^0$ & --- & $0.00051\pm 0.00016$ \\
Generic $D^+D^-$ (*) &$\to$& $K^0_S\pi^0$ & --- & $0.00038\pm 0.00012$ \\
$\bar K^{*0}\pi^+\pi^-$ &$\to$& $K^0_S\omega$ & 
        $2.3\pm 0.5$ & $0.065\pm 0.002$ \\
$K^{*-}\pi^+\pi^0$ &$\to$& $K^0_S\omega$ & 
        $1.00\pm 0.22$ & $0.135\pm 0.003$ \\
$K^{*-}\rho^+$ &$\to$& $K^0_S\omega$ & 
        $6.4\pm 2.5$ & $0.035\pm 0.001$ \\
$\bar K^{*0}\rho^0$ &$\to$& $K^0_S\omega$ & 
        $1.50\pm 0.33$ & $0.022\pm 0.004$ \\
$K^0_S\pi^+\pi^-\pi^0$ &$\to$& $K^0_S\omega$ & 
        $1.1\pm 1.1$ & $0.405\pm 0.423$ \\
$\pi^+\pi^-\pi^+\pi^-\pi^0$ &$\to$& $K^0_S\omega$ & 
        $0.41\pm 0.05$ & $0.006\pm 0.003$ \\
$K_1^-\pi^+$ &$\to$& $K^0_S\omega$ & 
        $1.12\pm 0.31$ & $0.038\pm 0.005$ \\
$\bar K_1^0\pi^0$ &$\to$& $K^0_S\omega$ & 
        $0.59\pm 0.02$ & $0.030\pm 0.010$ \\
$\eta\pi^0$ &$\to$& $K^0_L\pi^0$ & 
        $0.056\pm 0.014$ & $0.155\pm 0.001$ \\
$\pi^0\pi^0$ &$\to$& $K^0_L\pi^0$ & 
        $0.079\pm 0.008$ & $0.117\pm 0.001$ \\
\hline\hline
\end{tabular}
\end{center}
\end{table}

We also adjust the peaking background estimates for quantum
correlation effects.  The background branching fractions and efficiencies
mentioned above assume uncorrelated $D$ decay, but the background estimate
also depends on the type of decay of the other $D$.  For instance,
$K^0_L\pi^0$ signals have crossfeed background from $K^0_S\pi^0$ but only
for $K\pi$ and $CP$-even tags.  In $K^0_L\pi^0$ candidates opposite $CP$-odd
tags, $K^0_S\pi^0$ cannot contribute because it is also $CP$-odd.
For non-$CP$-eigenstate multi-body backgrounds, we assume equal $CP$-eigenstate
and flavored content, with the $CP$-eigenstate content equally divided
between $CP$-even and $CP$-odd.
Variations in these assumptions
give rise to systematic uncertainties, which are assessed in
Section~\ref{sec:results}.
In general, peaking backgrounds account for less than 1\% of the measured
yields, except in $K^0_S\omega$ modes (5--10\%), $K^0_L\pi^0$ modes (1--2\%),
and ST $K^0_S\pi^0\pi^0$ (3\%).


\section{External Measurements}\label{sec:externalMeas}

As discussed in Section~\ref{sec:formalism}, external estimates of uncorrelated
branching fractions can help determine $y$ and $\delta$.
We include measurements from Refs.~\cite{pdg06,pdg04} and from other
CLEO\nobreakdash-c analyses along with their full error matrix in the fit.
For the CLEO\nobreakdash-c analyses, we
account for statistical correlations with the yields measured in this analysis.
Tables~\ref{tab:externalBF},~\ref{tab:externalR2},
and~\ref{tab:externalYYPrime} shows the sources of these measurements, and
there is no overlap among the measurements in these Tables.

\subsection{External Branching Ratio Measurements}

Because our precision on $\cos\delta$ is currently limited by our knowledge
of the $CP$-eigenstate branching fractions, we include as many external
measurements of these branching fractions as possible.  
Except for the inclusive ${\cal B}(K^0_S\pi^0\pi^0)$ and
${\cal B}(K^0_L\pi^0)$, all $CP$-eigenstates
in this analysis have previous branching ratio measurements with respect to
other reference modes.

For $K^0_S\pi^0$, $K^0_S\eta$, and $K^0_S\omega$, we make use of the global
branching fraction fit performed in Ref.~\cite{pdg04}.
Because ${\cal B}(K^-\pi^+)$ is a free parameter in this fit and in
the current analysis, we also include Ref.~\cite{pdg04}'s fit result for this
mode in order to properly account for the correlations among these
branching fractions.  We do not use Ref.~\cite{pdg06} because this compilation
includes a CLEO\nobreakdash-c measurement of ${\cal B}(K^-\pi^+)$\cite{dhad56},
which is correlated with the measurements in this analysis.

Previous experiments have measured
$R_{KK}\equiv {\cal B}(K^-K^+)/{\cal B}(K^-\pi^+)$ and
$R_{\pi\pi}\equiv {\cal B}(\pi^-\pi^+)/{\cal B}(K^-\pi^+)$
simultaneously~\cite{pdg06}, so these two quantities are correlated
both statistically [via the common denominator, ${\cal B}(K^-\pi^+)$] and
systematically.  It is common for these experiments to also report a value of
$R_{KK}/R_{\pi\pi}$ in addition to $R_{KK}$ and $R_{\pi\pi}$ separately.
Using the dominant measurements of all three
quantities~\cite{kkpipiE791,kkpipiCLEO,kkpipiFOCUS,kkpipiCDF},
we compute a weighted average correlation coefficient of 0.30 between the
$R_{KK}$ and $R_{\pi\pi}$ values quoted in Ref~\cite{pdg06}.
For $R_{\pi\pi}$, we remove the CLEO\nobreakdash-c
measurement~\cite{pipiCLEOc} from the
average because it was based on the same dataset as this analysis.

The CLEO\nobreakdash-c $D\to K^0_{S/L}\pi^0$ analysis~\cite{ksklpi}, provides
additional information on ${\cal B}(K^0_L\pi^0)$ from the $K^-\pi^+\pi^0$ and
$K^-\pi^+\pi^-\pi^+$ tag modes, which are not used in this analysis.
Systematic correlations between the $K^0_L\pi^0$ yields in this analysis and
these additional ${\cal B}(K^0_L\pi^0)$ measurements is taken into
account.  We also include the statistical correlation between
${\cal B}(K^0_L\pi^0)$ and ${\cal B}(K^0_S\pi^0)$; knowledge of the
latter is required to correct for the effect of quantum correlations in
the former.

Table~\ref{tab:externalBF} summarizes the external branching fraction
inputs to the fit.

\begin{table}[htbp]
\begin{center}
\caption{External branching ratio measurements and their correlations
used in the data fit.  An asterisk (*) indicates removal of overlapping
CLEO-c measurements.}
\label{tab:externalBF}
\begin{tabular}{lcccccccc}
\hline\hline
Parameter & Value (\%) & 
        \multicolumn{7}{c}{Correlation Coefficients} \\
\hline
${\cal B}(K^-\pi^+)$    & $3.81\pm 0.09$~\cite{pdg04}
        & 1 & 0.08 & 0.06 & 0.04 & $-0.08$ & 0 & 0 \\
${\cal B}(K^0_S\pi^0)$  & $1.15\pm 0.12$~\cite{pdg04}
        &   & 1    & 0.58 & 0.14 & $-0.95$ & 0 & 0 \\
${\cal B}(K^0_S\eta)$   & $0.380\pm 0.060$~\cite{pdg04}
        &   &      & 1    & 0.10 & $-0.55$ & 0 & 0 \\
${\cal B}(K^0_S\omega)$ & $1.30\pm 0.30$~\cite{pdg04}
        &   &      &      & 1 & $-0.13$ & 0 & 0 \\
${\cal B}(K^0_L\pi^0)$ & $1.003\pm 0.083$*~\cite{ksklpi}
        &   &      &      &   &       1 & 0 & 0 \\
${\cal B}(K^-K^+)/{\cal B}(K^-\pi^+)$ & $10.10\pm 0.16$~\cite{pdg06}
        &   &   &      &      & & 1 & 0.30 \\
${\cal B}(\pi^-\pi^+)/{\cal B}(K^-\pi^+)$ & $3.588\pm 0.057$*~\cite{pdg06}
        &   &   &      &      & &   &  1  \\
\hline\hline
\end{tabular}
\end{center}
\end{table}

\subsection{External Mixing Measurements}

\subsubsection{Information on $r^2$ From Wrong-Sign
$D^0\to K^+\pi^-$ and $D^0\to K^{(*)+}\ell^-\nu_\ell$}

Studies of wrong-sign $D^0\to K^+\pi^-$ from other experiments provide
information on $r^2$, $y'$, and $R_{\rm M}$.  Without using $D^0$ decay times,
these analyses are sensitive to the time-integrated wrong-sign rate ratio:
\begin{eqnarray}
\label{eq:rws}
R_{\rm WS} &\equiv& \frac{\Gamma(D^0\to K^+\pi^-)}{\Gamma(\bar D^0\to K^+\pi^-)} \\
\label{eq:rws1}
        &=& r^2 + ry' + R_{\rm M} \\
\label{eq:rws2}
        &=& r^2 + \sqrt{r^2}y\cos\delta - \sqrt{r^2}x\sin\delta +
        \frac{x^2 + y^2}{2},
\end{eqnarray}
where Eq.~(\ref{eq:rws2}) explicitly shows the dependence on the fit parameters
$r^2$, $y$, $\cos\delta$, $x^2$, and $x\sin\delta$.
By also measuring the decay times of the same $D^0$ candidates, these
experiments can separate the three terms in Eq.~(\ref{eq:rws1}) because each
has a different proper time dependence:
\begin{equation}\label{eq:rwsTimeDep}
\frac{dN}{dt}\propto e^{-\Gamma t}\left[
r^2 + r y' \Gamma t + \frac{x'^2 + y'^2}{4} (\Gamma t)^2 \right].
\end{equation}

We use time-integrated $R_{\rm WS}$ measurements as a source of information
on $r^2$, which is otherwise poorly determined in our analysis, as shown in
Table~\ref{tab:noRWSRMResults}.  Constraining
$r^2$ leads directly to improved precision on $\cos\delta$ because it always
appears as $r\cos\delta$ in Table~\ref{tab:rates}.  
Extracting meaningful precision on $r^2$ from $R_{\rm WS}$ requires
information on $x\sin\delta$ and $x^2$.  When the external $y$ and $y'$
measurements described in Section~\ref{sec:externalYYPrime} are not included
in the fit, $x\sin\delta$ is poorly determined, so we take $x\sin\delta$ to
be zero.
For $x^2$, which is also poorly determined in our
analysis, we use $R_{\rm M}$ measurements from wrong-sign
$D^0\to K^{(*)+}\ell^-\nu_\ell$ rates.
Table~\ref{tab:externalR2} shows the input $R_{\rm WS}$ and semileptonic
$R_{\rm M}$ measurements, which are uncorrelated with all other measurements and
are entered directly into the fit.

\begin{table}[htbp]
\begin{center}
\caption{External mixing measurements used to constrain $r^2$.}
\label{tab:externalR2}
\begin{tabular}{lc}
\hline\hline
Parameter & Average (\%) \\
\hline
$R_{\rm WS}$ & $0.409\pm 0.022$~\cite{kpiE791,kpiFOCUS,kpiCDF} \\
$R_{\rm M}$ & $0.0173\pm 0.0387$~\cite{rmE791,rmCLEO,rmBelle,rmBABAR} \\
\hline\hline
\end{tabular}
\end{center}
\end{table}

\subsubsection{Measurements of $y$ and $y'$}\label{sec:externalYYPrime}

By combining external measurements of $y$ and $y'$ with the
CLEO\nobreakdash-c measurement of $\cos\delta$, the fitter can extract
$x\sin\delta$
via $x\sin\delta = y\cos\delta - y'$.  In doing so, it accounts for the
fact that $y'$ is determined simultaneously with $r^2$ and $R_{\rm M}$, which
are also functions of the fit parameters, and that $\cos\delta$ and its
uncertainty are correlated with $y$ (as discussed below in
Section~\ref{sec:results}).

External measurements of $y$ come from two sources:
comparison of $D^0$ decay times to $CP$ eigenstates $K^+K^-$ and
$\pi^+\pi^-$ with decay times to the flavor eigenstate $K^-\pi^+$, and
time-dependent Dalitz plot analysis of $D\to K^0_S\pi^+\pi^-$, which is
sensitive to both $x$ and $y$.  The latter $x$ and $y$ measurements are
essentially uncorrelated, but $x$ provides information on $y$ when combined
with measurements of $R_{\rm M}$.

For $y'$, we use three sets of ($CP$-conserving) fit results from CLEO, Belle,
and BABAR.  Averages of these fits are shown in
Table~\ref{tab:externalYYPrime}.  The fit covariance matrices have been
provided by the above collaborations and are also included in our analysis.

\begin{table}[htbp]
\begin{center}
\caption{External measurements of $y$ and $y'$ with associated
measurements of $r^2$ and $x'^2$.}
\label{tab:externalYYPrime}
\begin{tabular}{lc}
\hline\hline
Parameter & Value (\%) \\
\hline
$y$ & $0.662\pm 0.211$~\cite{kspipiBelle,kkpipiCLEO,ycpE791,ycpFOCUS,ycpBABAR,ycpBelle1,ycpBelle2,kspipiCLEO} \\
$x$ & $0.811\pm 0.334$~\cite{kspipiCLEO,kspipiBelle} \\
$r^2$ & $0.339\pm 0.012$~\cite{kpiBelle,kpiBABAR,kpiCLEO} \\
$y'$ & $0.34\pm 0.30$~\cite{kpiBelle,kpiBABAR,kpiCLEO} \\
$x'^2$ & $0.006\pm 0.018$~\cite{kpiBelle,kpiBABAR,kpiCLEO} \\
\hline\hline
\end{tabular}
\end{center}
\end{table}

We note that, if we assume $x\sin\delta=0$, the $y$ and $y'$ measurements in
Table~\ref{tab:externalYYPrime} by themselves provide an independent
measurement of $\cos\delta = ( y' + x\sin\delta ) / y = 0.52\pm 0.49$,
which is consistent with but less precise than
the value found independently with CLEO\nobreakdash-c data
(see Section~\ref{sec:results}).


\section{Systematic Uncertainties}\label{sec:systematics}

All uncertainties except statistical uncertainties on the measured yields are
considered to be systematic in origin.
Systematic uncertainties on the fit inputs are included directly in the
covariance matrix given to the fitter,
which propagates them to the fit parameters.  Uncorrelated yield uncertainties
are discussed above in Section~\ref{sec:selection}.
Below, we summarize the
sources of correlated uncertainty and the measurements to which they apply.

Final-state-dependent correlated systematic uncertainties, such as for
tracking, particle identification (PID), and $\pi^0$ and $K^0_S$ finding
efficiencies are
the dominant uncertainties for the branching fractions, but, as discussed in
Ref.~\cite{Asner:2005wf}, when no external measurements are included,
they cancel in the DCS and mixing parameters.  However, external measurements
are, in general, not subject to the same correlated effects as our measured
yields.  Thus, while these external measurements reduce the
statistical uncertainties on $y$ and $\cos\delta$, they also
introduce sensitivity to correlated uncertainties, and they increase the
systematic uncertainties somewhat.

In Table~\ref{tab:systematicsData}, we list the correlated systematic
uncertainties on reconstruction and PID efficiencies for final state
particles, determined as Refs.~\cite{dhad281,dshad} describe.
We also apply the efficiency corrections given in Refs.~\cite{dhad281,dshad}
with the following modifications.
For $\pi^0$ finding efficiency, we adjust the relative correction to $-3.3\%$
and inflate the systematic uncertainty to 4.0\% to account for our inclusion
of signal modes with more energetic $\pi^0$s than in Ref.~\cite{dhad281}.
For $\eta$ finding efficiency, the relative correction appropriate for our
selection criteria is $-6.5\%$ with an uncertainty of 4.0\%.
For $K^0_S$ finding efficiency, we include additional contributions of
0.3\% and 0.4\% for the flight significance and invariant mass requirements.
We also include uncertainties on peaking background branching fractions,
shown in Table~\ref{tab:backgroundSources}.
The signal and background efficiencies listed in this paper include the
corrections discussed above.

Table~\ref{tab:systematicsData2} shows mode-dependent uncertainties.
Most are assessed by relaxing selection requirements and noting the
resultant change in efficiency-corrected yield.  We compute each of these
uncertainties as the quadrature sum of the yield shift and its statistical
uncertainty, where the latter quantity accounts for the large
correlation between the two yields.

The effect of modeling of
the $K^0_S\pi^0\pi^0$ resonant substructure on the efficiency is determined
by reweighting the simulated events according to the substructure observed
in data and recomputing the efficiency.  We assign an uncertainty for the
modeling of multiple electron candidates by taking the difference between
multiple candidate rates in simulated events and data.  The efficiency of
the extra track, $\pi^0$, and $\eta$ vetos in $K^0_L\pi^0$ modes is studied
by comparing the frequency of these extra particles
between simulated events and data.
For semileptonic decays, we extrapolate the observed electron momentum
spectrum below 200 MeV/$c$ based on the simulation.  By sampling different
form factor models used in the simulation, we estimate a relative
uncertainty of 25\% on this extrapolation.

In our simulation, final state radiation (FSR) is generated by
PHOTOS~\cite{photos}.  We measure efficiencies with and without FSR generation
and assign 30\% of the difference as a systematic uncertainty~\cite{dhad281}.
We also include a contribution
corresponding to the change in efficiency when PHOTOS does and does not model
interference among the radiated photons.

Unlike FSR, initial state radiation (ISR) coherently shifts both $M$
values upward in a signal DT.  So, mismodeling of ISR would affect
ST and DT efficiencies by the same fraction.  Based on a study of excluding
events in the ISR region ($M > 1.87$ GeV/$c^2$), 
we assign a conservative 0.5\% uncertainty to all yields.

\begin{table}[htbp]
\begin{center}
\caption{Correlated, fractional efficiency systematic uncertainties and
the schemes for applying them in the data branching fraction fit.}
\label{tab:systematicsData}
\begin{tabular}{lcc}
\hline\hline
Source & Uncertainty (\%) & Scheme \\ \hline
Track finding & 0.3 & per track \\
$K^\pm$ hadronic interactions & 0.6 & per $K^\pm$ \\
$K^0_S$ finding & 1.9 & per $K^0_S$ \\
$\pi^0$ finding & 4.0 & per $\pi^0$ \\
$\eta$ finding & 4.0 & per $\eta$ \\
$dE/dx$ and RICH & 0.3 & per $\pi^\pm$ PID cut \\
$dE/dx$ and RICH & 0.3 & per $K^\pm$ PID cut \\
EID & 1.0 & per $e^\pm$ \\
\hline\hline
\end{tabular}
\end{center}
\end{table}

\begin{table}[htbp]
\begin{center}
\caption{Correlated, mode-dependent fractional systematic uncertainties
in percent for STs.  An asterisk (*) marks those uncertainties that
are correlated among modes.  The schemes by which these uncertainties are
applied to DTs are also given, along with the fractional DT
uncertainty, $\lambda_{\rm DT}$, on mode $\{A, B\}$ for ST uncertainties of
$\alpha$ on mode $A$ and $\beta$ on mode $B$.}
\label{tab:systematicsData2}
\begin{tabular}{lcccccl}
\hline\hline
   & $\Delta E$ & ISR* & FSR* & Lepton Veto* & \multicolumn{2}{l}{Other} \\
\hline
$K^\mp\pi^\pm$ & 0.5 & 0.5 & 1.2 & 0.5 \\
$K^+K^-$ & 0.9 & 0.5 & 0.8 & 0.4 & 0.5 & $K^\pm$ $\cos\theta$ cut \\
$\pi^+\pi^-$ & 1.9 & 0.5 & 1.7 & 3.2 \\
$K^0_S\pi^0\pi^0$ & 2.6 & 0.5 & & & 1.5 & $K^0_S$ daughter PID \\
	& & & & & 0.7 & resonant substructure \\
$K^0_S\pi^0$ & 0.9 & 0.5 \\
$K^0_S\eta$ & 5.5 & 0.5 & & & 0.3 & $\eta$ mass cut \\
	& & & & & 0.7 & ${\cal B}(\eta\to\gamma\gamma)$~\cite{pdg06} \\
$K^0_S\omega$ & 1.2 & 0.5 & 0.8 & & 1.4 & $\omega$ mass cut \\
	& & & & & 0.8 & ${\cal B}(\omega\to\pi^+\pi^-\pi^0)$~\cite{pdg06} \\
$Xe\nu_e$ & & 0.5 & 0.3 & & 2.0 & spectrum extrapolation \\
	& & & & & 0.7 & multiple $e^\pm$ candidates \\
$K^0_L\pi^0$ & & 0.5 & & & 0.7 & background subtraction \\
	& & & & & 0.3 & extra track veto \\
	& & & & & 1.4 & signal shape \\
	& & & & & 1.6 & extra $\pi^0$ veto \\
	& & & & & 0.5 & $\eta$ veto \\
\hline
Scheme & per $D$ & per yield & per $D$ & per ST & \multicolumn{2}{l}{per $D$}\\
$\lambda_{\rm DT}$ & $\sqrt{\alpha^2+\beta^2}$ & $(\alpha+\beta)/2$
        & $\alpha+\beta$ & 0 & $\sqrt{\alpha^2+\beta^2}$ \\
\hline\hline
\end{tabular}
\end{center}
\end{table}


\section{Fit Results}\label{sec:results}

We tested the analysis technique with the full simulated sample of
uncorrelated $D^0\bar D^0$ decays that was filtered to leave a 40\% subset
of events that mimic the effect of quantum correlations.  The effective
luminosity of this subset is 15 times that of the data sample.
The fit results showed satisfactory agreement with the input values,
taking into account the statistical correlation between the signal
efficiencies (measured in the full simulated sample) and the signal yields
(measured in the 40\% subset).

In Table~\ref{tab:noRWSRMResults}, we first show the results of a data fit that
includes external branching fraction measurements from
Table~\ref{tab:externalBF}, but not the external
$R_{\rm WS}$ and $R_{\rm M}$ measurements in Table~\ref{tab:externalR2}.
The corresponding correlation matrix for the fit parameters is given in
Table~\ref{tab:noRWSRMCorrs}.
Because $r^2$ has a large uncertainty and a negative central value, we
cannot extract $\cos\delta$; instead, we quote $r\cos\delta$.  Also, we fit
for $R_{\rm M}$ instead of $x^2$.

\begin{table}[htbp]
\begin{center}
\caption{Results from the fit with external inputs from
Table~\ref{tab:externalBF}, but not Tables~\ref{tab:externalR2}
or~\ref{tab:externalYYPrime}.
Uncertainties are statistical and systematic, respectively.
Charge-averaged $D^0$ branching fractions are denoted by final state.}
\label{tab:noRWSRMResults}
\begin{tabular}{lc}
\hline\hline
Parameter & Fitted Value \\
\hline
${\cal N}$ $(10^6)$ &
	$1.062\pm 0.024\pm 0.011$ \\
$y$ $(10^{-3})$ &
	$-52\pm 60\pm 17$ \\
$r^2$ $(10^{-3})$ &
	$-24\pm 16\pm 12$ \\
$r\cos\delta$ &
	$0.089\pm 0.036\pm 0.009$ \\
$R_{\rm M}$ $(10^{-3})$ &
	$2.0\pm 1.2\pm 1.2$ \\
$rx\sin\delta$ $(10^{-3})$ &
        0 (fixed) \\
${\cal B}(K^-\pi^+)$ (\%) &
	$3.81\pm 0.05\pm 0.06$ \\
${\cal B}(K^-K^+)$ $(10^{-3})$ &
	$3.86\pm 0.06\pm 0.06$ \\
${\cal B}(\pi^-\pi^+)$ $(10^{-3})$ &
	$1.37\pm 0.02\pm 0.03$ \\
${\cal B}(K^0_S\pi^0\pi^0)$ $(10^{-3})$ &
	$8.27\pm 0.45\pm 0.41$ \\
${\cal B}(K^0_S\pi^0$) (\%) &
	$1.13\pm 0.03\pm 0.03$ \\
${\cal B}(K^0_S\eta)$ $(10^{-3})$ &
	$4.36\pm 0.15\pm 0.27$ \\
${\cal B}(K^0_S\omega)$ (\%) &
	$1.10\pm 0.04\pm 0.05$ \\
${\cal B}(X^- e^+\nu_e)$ (\%) &
	$6.35\pm 0.19\pm 0.18$ \\
${\cal B}(K^0_L\pi^0$) (\%) &
	$1.02\pm 0.03\pm 0.02$ \\
\hline
$\chi^2_{\rm fit}$/ndof &
	26.1/44 \\
\hline\hline
\end{tabular}
\end{center}
\end{table}

\begin{table}[htbp]
\begin{center}
\caption{Correlation coefficients (\%) for the fit in
Table~\ref{tab:noRWSRMResults} using both statistical and
systematic uncertainties.}
\label{tab:noRWSRMCorrs}
\begin{tabular}{c|rrrrrrrrrrrrr}
\hline\hline
& $y$ & $r^2$ & $r\cos\delta$ & $R_{\rm M}$ & $K^-\pi^+$ & $K^-K^+$ &
	$\pi^-\pi^+$ &
        $K^0_S\pi^0\pi^0$ & $K^0_S\pi^0$ & $K^0_S\eta$ & 
        $K^0_S\omega$ & $X^- e^+\nu_e$ & $K^0_L\pi^0$ \\
\hline
${\cal N}$ &
	$-13$ & $-37$ & $-3$ & $5$ & $-44$ & $-60$ & $-46$ & $-17$ & $-46$ &
		$-27$ & $-28$ & $-17$ & $37$ \\
$y$ &
	      & $-32$ & $-83$ & $-6$ & $-23$ & $12$ & $7$ & $-0$ & $-4$ &
		$2$ & $-3$ & $30$ & $7$ \\
$r^2$ &
	      &     & $39$ & $1$ & $-16$ & $7$ & $-6$ & $8$ & $23$ &
		$11$ & $20$ & $29$ & $-20$ \\
$r\cos\delta$ &
	      &     &      & $6$ & $23$ & $-3$ & $-1$ & $12$ & $21$ &
		$4$ & $13$ & $-21$ & $-22$ \\
$R_{\rm M}$ &
	      &     &      &       & $-9$ & $-3$ & $-3$ & $-0$ & $-1$ &
		$-1$ & $-1$ & $-2$ & $1$ \\
$K^-\pi^+$ &
	      &     &      &       &       & $72$ & $74$ & $10$ & $27$ &
		$14$ & $21$ & $-16$ & $-22$ \\
$K^-K^+$ &
	      &     &      &       &       &      & $73$ & $12$ & $31$ &
		$18$ & $24$ & $9$ & $-25$ \\
$\pi^-\pi^+$ &
	      &     &      &       &       &      &      & $9$ & $25$ &
		$14$ & $20$ & $-0$ & $-19$ \\
$K^0_S\pi^0\pi^0$ &
	      &     &      &       &       &      &      &      & $64$ &
		$15$ & $46$ & $3$ & $-37$ \\
$K^0_S\pi^0$ &
	      &     &      &       &       &      &      &      &       &
		$31$ & $59$ & $8$ & $-66$ \\
$K^0_S\eta$ &
	      &     &      &       &       &      &      &      &       &
		     & $20$ & $5$ & $-22$ \\
$K^0_S\omega$ &
	      &     &      &       &       &      &      &      &       &
		     &      & $9$ & $-38$ \\
$X^- e^+\nu_e$ &
	      &     &      &       &       &      &      &      &       &
		     &      &       & $-5$ \\
\hline\hline
\end{tabular}
\end{center}
\end{table}

In order to control the uncertainty on $\cos\delta$, we include in our
standard fit both external branching fractions from
Table~\ref{tab:externalBF} as well as external $R_{\rm WS}$ and $R_{\rm M}$
measurements from Table~\ref{tab:externalR2}.  
In this fit, shown in Table~\ref{tab:dataResults},
we obtain a first measurement of $\cos\delta$, consistent with being at
the boundary of the physical region and with a precision that is dominated by
the CLEO\nobreakdash-c $\{K\pi, S_\pm\}$ yield measurements.  
Our value of $y$ is consistent with the world average
of $0.00662\pm 0.00211$ (see Table~\ref{tab:externalYYPrime}).
For this standard fit, we assume $x\sin\delta=0$, and the associated systematic
uncertainty is $\pm 0.03$ for $\cos\delta$ and negligible for all other
parameters.
The correlation matrix for this standard
fit is shown in Table~\ref{tab:corrs1}.
Our branching fraction results do not supersede other CLEO\nobreakdash-c
measurements.

\begin{table}[htb]
\caption{Results from the standard fit (with inputs from
Tables~\ref{tab:externalBF} and~\ref{tab:externalR2}) and the
extended fit (with inputs from
Tables~\ref{tab:externalBF}, \ref{tab:externalR2},
and~\ref{tab:externalYYPrime}).  Uncertainties
are statistical and
systematic, respectively.  Charge-averaged $D^0$ branching fractions are
denoted by final state.}
\label{tab:dataResults}
\begin{tabular}{lcc}
\hline\hline
Parameter & Standard Fit & Extended Fit \\
\hline
${\cal N}$ $(10^6)$ &
	$1.042\pm 0.021\pm 0.010$ &
        $1.042\pm 0.021\pm 0.010$ \\
$y$ $(10^{-3})$ &
	$-45\pm 59\pm 15$ &
        $6.5\pm 0.2\pm 2.1$ \\
$r^2$ $(10^{-3})$ &
        $8.0\pm 6.8\pm 1.9$ &
        $3.44\pm 0.01\pm 0.09$ \\
$\cos\delta$ &
        $1.03\pm 0.19\pm 0.06$ &
        $1.10\pm 0.35\pm 0.07$ \\
$x^2$ $(10^{-3})$ &
        $-1.5\pm 3.6\pm 4.2$ &
        $0.06\pm 0.01\pm 0.05$ \\
$x\sin\delta$ $(10^{-3})$ &
	0 (fixed) &
	$4.4\pm 2.4\pm 2.9$ \\
$K^-\pi^+$ (\%) &
        $3.78\pm 0.05\pm 0.05$ &
        $3.78\pm 0.05\pm 0.05$ \\
$K^-K^+$ $(10^{-3})$ &
        $3.87\pm 0.06\pm 0.06$ &
        $3.88\pm 0.06\pm 0.06$ \\
$\pi^-\pi^+$ $(10^{-3})$ &
        $1.36\pm 0.02\pm 0.03$ &
        $1.36\pm 0.02\pm 0.03$ \\
$K^0_S\pi^0\pi^0$ $(10^{-3})$ &
        $8.34\pm 0.45\pm 0.42$ &
        $8.35\pm 0.44\pm 0.42$ \\
$K^0_S\pi^0$ (\%) &
        $1.14\pm 0.03\pm 0.03$ &
        $1.14\pm 0.03\pm 0.03$ \\
$K^0_S\eta$ $(10^{-3})$ &
        $4.42\pm 0.15\pm 0.28$ &
        $4.42\pm 0.15\pm 0.28$ \\
$K^0_S\omega$ (\%) &
        $1.12\pm 0.04\pm 0.05$ &
        $1.12\pm 0.04\pm 0.05$ \\
$X^- e^+\nu_e$ (\%) &
        $6.54\pm 0.17\pm 0.17$ &
        $6.59\pm 0.16\pm 0.16$ \\
$K^0_L\pi^0$ (\%) &
        $1.01\pm 0.03\pm 0.02$ &
        $1.01\pm 0.03\pm 0.02$ \\
\hline
$\chi^2_{\rm fit}$/ndof &
        30.1/46 &
        55.3/57 \\
\hline\hline
\end{tabular}
\end{table}

\begin{table}[htbp]
\begin{center}
\caption{Correlation coefficients (\%) for the standard fit in
Table~\ref{tab:dataResults} using both statistical and
systematic uncertainties.}
\label{tab:corrs1}
\begin{tabular}{c|rrrrrrrrrrrrr}
\hline\hline
& $y$ & $r^2$ & $\cos\delta$ & $x^2$ & $K^-\pi^+$ & $K^-K^+$ & $\pi^-\pi^+$ &
	$K^0_S\pi^0\pi^0$ & $K^0_S\pi^0$ & $K^0_S\eta$ & 
        $K^0_S\omega$ & $X^- e^+\nu_e$ & $K^0_L\pi^0$ \\
\hline
${\cal N}$ &
	$-11$ & $8$ & $-22$ & $-10$ & $-65$ & $-61$ & $-53$ & $-15$ & $-42$ &
		$-24$ & $-23$ & $2$ & $33$ \\
$y$ &
	      & $-99$ & $49$ & $99$ & $-20$ & $10$ & $6$ & $-0$ & $-6$ &
		$1$ & $-4$ & $30$ & $7$ \\
$r^2$ &
	      &     & $-38$ & $-99$ & $21$ & $-8$ & $-5$ & $3$ & $9$ &
		$0$ & $6$ & $-30$ & $-11$ \\
$\cos\delta$ &
	      &     &      & $50$ & $2$ & $15$ & $12$ & $17$ & $22$ &
		$8$ & $13$ & $13$ & $-22$ \\
$x^2$ &
	      &     &      &       & $-20$ & $10$ & $6$ & $-0$ & $-6$ &
		$1$ & $-4$ & $30$ & $7$ \\
$K^-\pi^+$ &
	      &     &      &       &       & $80$ & $77$ & $12$ & $35$ &
		$19$ & $28$ & $-5$ & $-28$ \\
$K^-K^+$ &
	      &     &      &       &       &      & $74$ & $11$ & $30$ &
		$19$ & $28$ & $-6$ & $-28$ \\
$\pi^-\pi^+$ &
	      &     &      &       &       &      &      & $10$ & $26$ &
		$15$ & $22$ & $1$ & $-21$ \\
$K^0_S\pi^0\pi^0$ &
	      &     &      &       &       &      &      &      & $64$ &
		$14$ & $46$ & $0$ & $-36$ \\
$K^0_S\pi^0$ &
	      &     &      &       &       &      &      &      &       &
		$29$ & $57$ & $-2$ & $-65$ \\
$K^0_S\eta$ &
	      &     &      &       &       &      &      &      &       &
		     & $18$ & $-0$ & $-21$ \\
$K^0_S\omega$ &
	      &     &      &       &       &      &      &      &       &
		     &      & $1$ & $-36$ \\
$X^- e^+\nu_e$ &
	      &     &      &       &       &      &      &      &       &
		     &      &       & $2$ \\
\hline\hline
\end{tabular}
\end{center}
\end{table}

\begin{table}[htbp]
\begin{center}
\caption{Correlation coefficients (\%) for the extended fit in
Table~\ref{tab:dataResults} using both statistical and
systematic uncertainties.}
\label{tab:corrs2}
\begin{tabular}{c|rrrrrrrrrrrrrr}
\hline\hline
& $y$ & $r^2$ & $\cos\delta$ & $x^2$ & $x\sin\delta$ &
	$K^-\pi^+$ & $K^-K^+$ & $\pi^-\pi^+$ &
	$K^0_S\pi^0\pi^0$ & $K^0_S\pi^0$ & $K^0_S\eta$ & 
        $K^0_S\omega$ & $X^- e^+\nu_e$ & $K^0_L\pi^0$ \\
\hline
${\cal N}$ &
	$0$ & $-0$ & $-19$ & $0$ & $-12$ & $-69$ & $-61$ & $-53$ & $-15$ &
		$-43$ & $-24$ & $-23$ & $5$ & $34$ \\
$y$ &
	      & $13$ & $-6$ & $-6$ & $69$ & $0$ & $0$ & $-0$ & $0$ & $-0$ &
		$-0$ & $-0$ & $1$ & $0$ \\
$r^2$ &
	      &     & $-5$ & $26$ & $37$ & $-0$ & $0$ & $-0$ & $0$ & $0$ &
		$0$ & $0$ & $0$ & $-0$ \\
$\cos\delta$ &
	      &     &      & $-1$ & $56$ & $14$ & $12$ & $10$ & $20$ & $29$ &
		$9$ & $17$ & $1$ & $-30$ \\
$x^2$ &
	      &     &      &      & $19$ & $-0$ & $-0$ & $-0$ & $-0$ & $-0$ &
		$-0$ & $-0$ & $-0$ & $0$ \\
$x\sin\delta$ &
	      &     &      &      &      & $8$ & $8$ & $6$ & $13$ & $18$ &
		$5$ & $11$ & $1$ & $-18$ \\
$K^-\pi^+$ &
	      &     &      &       &     &     & $84$ & $80$ & $12$ & $34$ &
		$19$ & $28$ & $1$ & $-27$ \\
$K^-K^+$ &
	      &     &      &       &     &     &      & $74$ & $11$ & $31$ &
		$17$ & $23$ & $1$ & $-24$ \\
$\pi^-\pi^+$ &
	      &     &      &       &     &     &      &      & $10$ & $27$ &
		$15$ & $22$ & $-0$ & $-21$ \\
$K^0_S\pi^0\pi^0$ &
	      &     &      &       &     &     &      &      &      & $64$ &
		$14$ & $46$ & $1$ & $-36$ \\
$K^0_S\pi^0$ &
	      &     &      &       &     &     &      &      &      &      &
		$29$ & $57$ & $1$ & $-65$ \\
$K^0_S\eta$ &
	      &     &      &       &     &     &      &      &      &      &
		     & $18$ & $-0$ & $-21$ \\
$K^0_S\omega$ &
	      &     &      &       &     &     &      &      &      &      &
		     &      & $3$ & $-36$ \\
$X^- e^+\nu_e$ &
	      &     &      &       &     &     &      &      &      &      &
		     &      &       & $-1$ \\
\hline\hline
\end{tabular}
\end{center}
\end{table}

\begin{figure}[htb]
\includegraphics*[width=\linewidth]{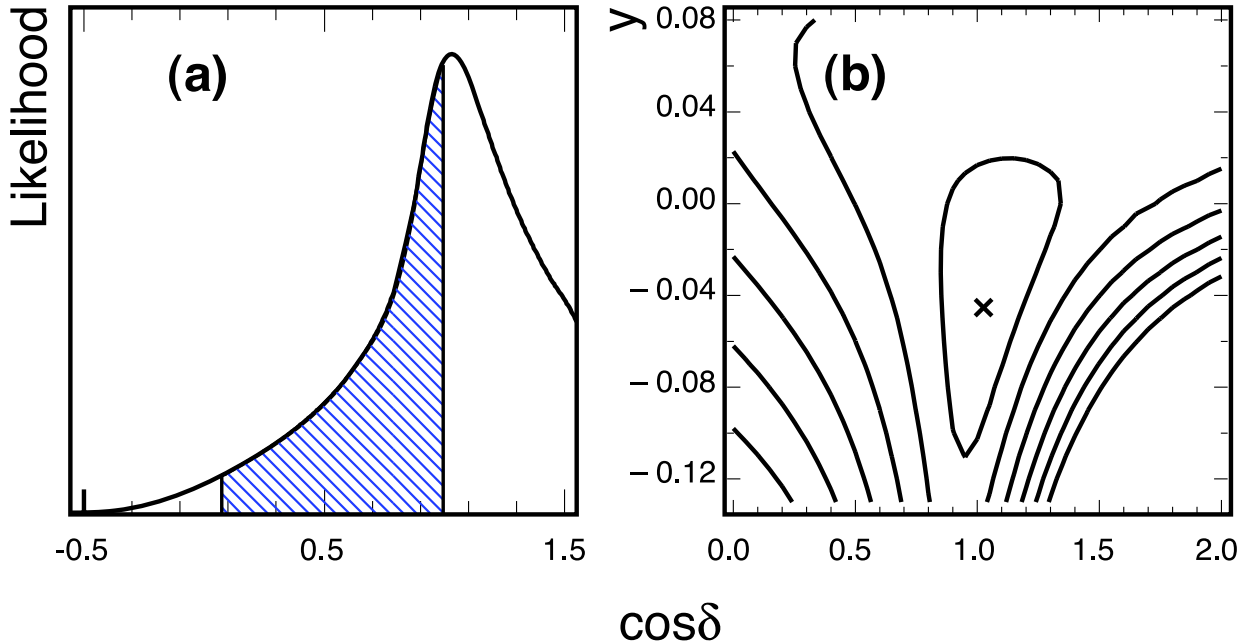}
\caption{Standard fit likelihood including both statistical and systematic
uncertainties for $\cos\delta$ (a), and simultaneous likelihood for
$\cos\delta$ and $y$ (b) shown as contours in increments of $1\sigma$, where
$\sigma=\sqrt{\Delta\chi^2}$.  
The hatched region contains 95\% of the area in the physical region.}
\label{fig:standardFitLikelihoods}
\end{figure}

The likelihood curve for $\cos\delta$ is obtained by repeating this fit at
fixed values of $\cos\delta$ and recording the change in minimum $\chi^2$.
We then compute ${\cal L}=e^{-(\chi^2-\chi^2_{\rm min})/2}$, which is shown
in Fig.~\ref{fig:standardFitLikelihoods}a. It is highly non-Gaussian, so we
assign asymmetric uncertainties by finding
the values of $\cos\delta$ where $\Delta\chi^2=1$ to obtain
$\cos\delta = 1.03^{+0.31}_{-0.17}\pm 0.06$.
For values of $\left|\cos\delta\right| < 1$, we also compute
${\cal L}$ as a function of $|\delta|$, and we integrate these
curves within the physical region to obtain 95\% confidence level limits
of $\cos\delta > 0.07$ and $|\delta| < 75^\circ$.

The asymmetric uncertainties on $\cos\delta$ quoted above still do not fully
capture the non-linearity of the likelihood.  This non-linearity stems from the
use of $r^2$ to convert $r\cos\delta$ into $\cos\delta$, which causes the
uncertainty on $\cos\delta$ to scale roughly like $1/r$.  Because $r^2$ is
obtained from $R_{\rm WS}$ via Eq.~(\ref{eq:rws2}), an upward shift in $y$
lowers the derived value of $r^2$ (for positive $r\cos\delta$).  As a
result, the uncertainty on $\cos\delta$ increases with more positive $y$,
as shown in Fig.~\ref{fig:standardFitLikelihoods}b.
A second, smaller effect is that the sign of the
correlation between $r^2$ and $r\cos\delta$ is given by the sign of $y$.
When $y$ is positive, $r^2$ and $r\cos\delta$ are anti-correlated, which
tends to inflate the uncertainty on $\cos\delta$.

\begin{figure}[htb]
\includegraphics*[width=\linewidth]{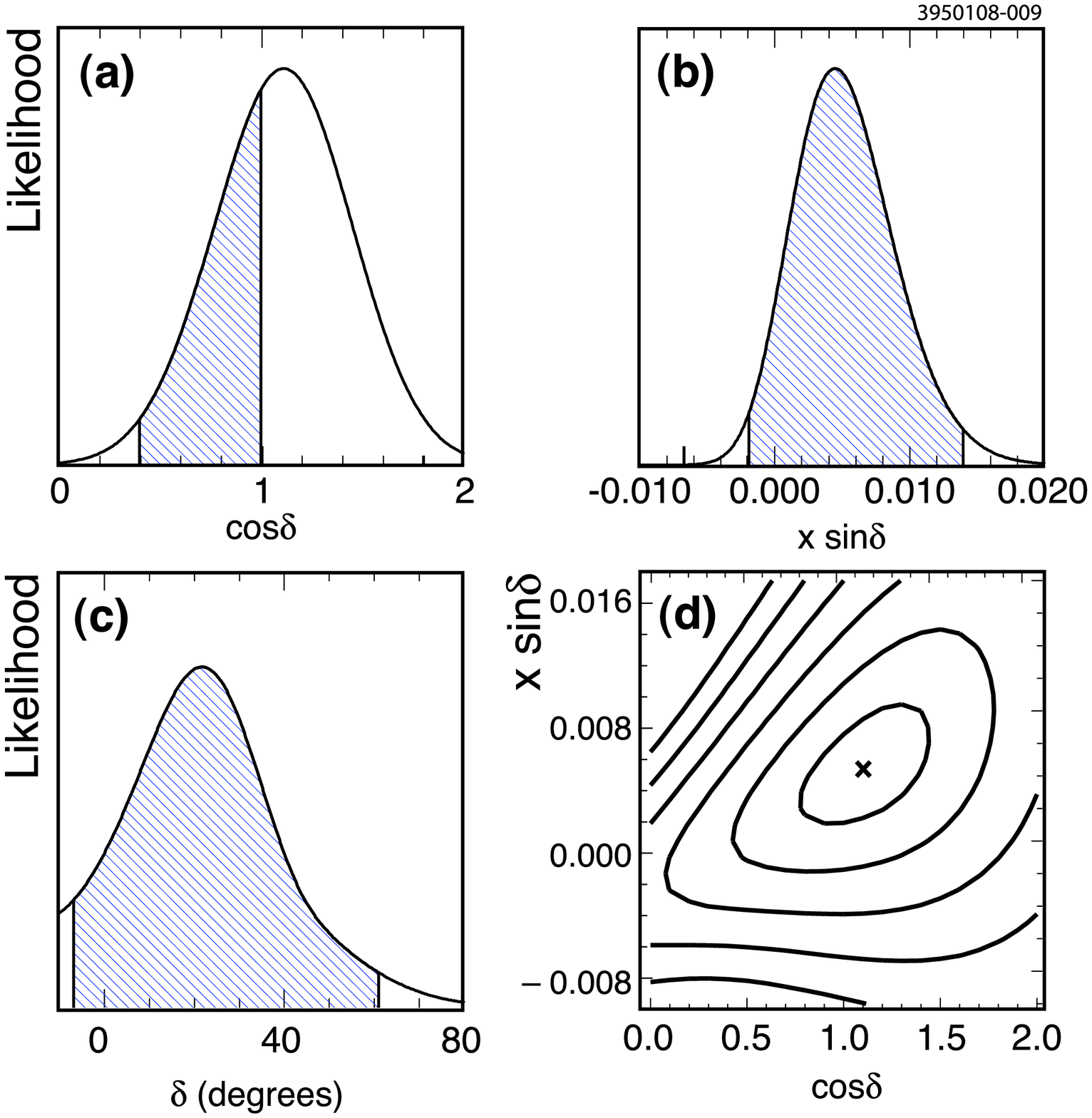}
\caption{Extended fit likelihood including both statistical and systematic
uncertainties for $\cos\delta$ (a), $x\sin\delta$ (b), $\delta$ (c), and
simultaneous likelihood for $\cos\delta$ and $x\sin\delta$ (d) shown as
contours in increments of $1\sigma$, where $\sigma=\sqrt{\Delta\chi^2}$.
The hatched
regions contain 95\% of the area in the physical regions.  For $\delta$,
the fit fails to converge beyond the limits of the plot.}
\label{fig:extendedFitLikelihoods}
\end{figure}

We also perform an extended fit that determines $x\sin\delta$ by
including previous measurements of $y$ and $y'$.  Table~\ref{tab:dataResults}
shows the results of this fit, which incorporates all external
measurements, from Tables~\ref{tab:externalBF}, \ref{tab:externalR2},
and~\ref{tab:externalYYPrime}.  The correlation matrix for the extended fit
is given in Table~\ref{tab:corrs2}.
The resultant value of $y$ includes the
CLEO\nobreakdash-c measurement from the standard fit, but the precision is
dominated by the external $y$ measurements.
The overall uncertainty on $\cos\delta$ increases to $\pm 0.36$ because of
the $R_{\rm WS}$ measurement, as discussed above.
However, unlike the standard fit, the likelihood for $\cos\delta$ shown in
Fig.~\ref{fig:extendedFitLikelihoods}a is nearly Gaussian.
For $x\sin\delta$, we assign asymmetric uncertainties resulting in
$x\sin\delta = (4.4^{+2.7}_{-1.8}\pm 2.9)\times 10^{-3}$.
By repeating the fit at various simultaneously fixed values of $\cos\delta$
and $\sin\delta$, we also determine
$\delta = (22^{+11}_{-12}$$^{+9}_{-11})^\circ$.
The corresponding 95\% confidence level intervals within the physical region
are $\cos\delta > 0.39$, $x\sin\delta\in [0.002, 0.014]$, and
$\delta\in [-7^\circ, +61^\circ]$.
Performing this extended fit with $y$, $x^2$, and $x\sin\delta$ fixed to zero
results in a change in $\chi^2$ of 25.1, or a significance of $5.0\sigma$.

In the standard fit, the large correlation between $y$ and $r^2$ originates
in the use of the $R_{\rm WS}$ measurement in
Table~\ref{tab:externalR2} to provide information on $r^2$.
Using Eq.~(\ref{eq:rws2}), one sees that the precision on $r^2$ is driven by
the precision on $y$.  Hence, for positive $r\cos\delta$, $r^2$ and $y$ are
negatively correlated.  Consequently, the $\cos\delta$ value obtained by
dividing $r\cos\delta$ by $\sqrt{r^2}$ is positively correlated with $y$.
In the extended fit, these correlations are broken by the precise external
measurements of $y$ in Table~\ref{tab:externalYYPrime}, and the weak negative
correlation between $\cos\delta$ and $y$ results from Eqs.~(\ref{eq:rz1})
and~(\ref{eq:rz1a}).

We assess the fit's sensitivity to the assumed quantum correlation parameters
in the peaking background subtractions by varying these parameters over their
full allowed ranges.  We find excursions of less than $1\times 10^{-4}$
in $y$ and $0.003$ in $\cos\delta$ for both the standard and extended fits.
We assign a systematic uncertainty, included in
Table~\ref{tab:dataResults} given by the size of these shifts.

\subsection{Analysis of Input Information}\label{sec:statPower}

In this Section, we probe the power of the individual fit inputs
or groups of fit inputs by removing them from the standard fit one by one.
We define the weight (or information) that each input contributes to fit
parameter $\lambda$ to be the fractional change in $1/\sigma_\lambda^2$
when it is removed from the fit:
\begin{equation}
I_\lambda\equiv \frac{1/\sigma_\lambda^2({\rm new}) -
1/\sigma_\lambda^2({\rm standard})}{1/\sigma_\lambda^2({\rm standard})}.
\end{equation}
Because of the aforementioned non-linearities, removing an input from the fit
can result in a $\sigma_\lambda$ that is {\it smaller} than the standard value,
leading to negative $I_\lambda$.

Table~\ref{tab:information} shows $I_\lambda$ for $\lambda = y$ and
$\cos\delta$.
When $I_\lambda > 0$, we also calculate the significance of the shift to be
${\cal S}_\lambda\equiv\Delta\lambda/\sqrt{\sigma_\lambda^2({\rm standard})-\sigma_\lambda^2({\rm new})
}$, where the minus sign accounts for the correlation between the two values
of $\lambda$.

As expected from Eqs.~(\ref{eq:y1}), (\ref{eq:y1a}), and~(\ref{eq:y2}),
information about $y$ comes primarily from the $\{S_\pm, e\}$ DT yields,
which supply 91\% of the statistical power. In addition, the $\{K\pi, e\}$
yields provide the normalizing semileptonic branching fraction, and
$S_\pm$ ST yields provide ${\cal N}{\cal B}_{S_\pm}$.
Because $y$ is determined from ratios of these quantities, a meaningful
measurement depends on the simultaneous presence of various combinations
of inputs.  For instance, removing the $S_{\pm}$ ST yields would also
reduce the power of the $\{S_\pm, e\}$ DT yields because, then,
Eq.~(\ref{eq:y1}) cannot be used.  Table~\ref{tab:information} does not
account for such double-counting of information; it simply shows
$I_\lambda$ as each individual input is removed from the fit.  For this
reason, the sum of all $I_y$ exceeds 100\%.

This effect is even more pronounced with $\cos\delta$.  As demonstrated
in Eqs.~(\ref{eq:rz1}) and~(\ref{eq:rz2}), almost all the information
about $r\cos\delta$ is contained in the $\{K\pi, S_\pm\}$ DT yields,
but only when they are combined with the ${\cal B}_{S_\pm}$ estimates
provided by
the $S_\pm$ ST yields.  Furthermore, in order to obtain $\cos\delta$
from $r\cos\delta$, we rely on the inputs from Table~\ref{tab:externalR2},
which therefore have $I_{\cos\delta}=1$.
If either $\{K\pi, S_+\}$ or $\{K\pi, S_-\}$ yields are removed, then the
asymmetries in Eqs.~(\ref{eq:rz1}) and~(\ref{eq:rz2}) cannot be formed.
However, $r\cos\delta$ can still be obtained with knowledge of
${\cal N}{\cal B}_{K^-\pi^+}{\cal B}_{S_\pm}$, which comes from other
inputs in the fit.

A striking example of non-linearities in the fit occurs when
$\{S_-, e^\pm\}$ double tags are removed.  Doing so causes $y$ to fluctuate
downward, which, in turn, {\it lowers} the uncertainty on $\cos\delta$,
as discussed above.  As a result, $\{S_-, e^\pm\}$
double tags appear to have large {\it negative} information content.

In performing this exercise, we also test for anomalous inputs that bias
the fit results with undue weight.  As Table~\ref{tab:information} shows,
we find no evidence of instability from our choice of fit inputs.

\begin{table}[htbp]
\begin{center}
\caption{Effect of removing inputs from the standard fit.
For each input and fit parameter combination, we
report the significance of the shift in central value ${\cal S}_\lambda$
and the fractional information content $I_\lambda$.  In cases
where $I_\lambda$ is negative, we do not report the shift significance.}
\label{tab:information}
\begin{tabular}{ccccc}
\hline\hline
Removed Input & ${\cal S}_y$ & $I_y$ & ${\cal S}_{\cos\delta}$ &
	$I_{\cos\delta}$ \\
\hline
\multicolumn{5}{c}{Single Tag Yields} \\
$K^-\pi^+$ & $+0.1\sigma$ & 0\% &
	$-0.0\sigma$ & 0\% \\
$K^+\pi^-$ & $+0.2\sigma$ & 0\% &
	$-0.0\sigma$ & $1\%$ \\
$K^+K^-$ & $+1.1\sigma$ & 6\% &
	$+0.4\sigma$ & 41\% \\
$\pi^+\pi^-$ & $+0.6\sigma$ & 1\% &
	$+0.2\sigma$ & 11\% \\
$K^0_S\pi^0\pi^0$ & $+0.5$ & $2\%$ &
	$+0.4\sigma$ & 23\% \\
$K^0_S\pi^0$ & $+1.1\sigma$ & 14\% &
	$+0.5\sigma$ & 72\% \\
$K^0_S\eta$ & $-2.8\sigma$ & 0\% &
	--- & $-31\%$ \\
$K^0_S\omega$ & $-2.5\sigma$ & 2\% &
	--- & $-46\%$ \\
\hline
\multicolumn{5}{c}{Double Tag Yields} \\
$K^\pm\pi^\mp, K^\pm\pi^\mp$ & $+0.1\sigma$ & $1\%$ &
	$+0.1\sigma$ & 8\% \\
$K^-\pi^+, K^+\pi^-$ & --- & $-0\%$ &
	--- & $-2\%$ \\
$K\pi, S_+$ & --- & $-0\%$ &
	$-0.0\sigma$ & 48\% \\
$K\pi, S_-$ & $-0.1\sigma$ & 0\% &
	$+0.0\sigma$ & 38\% \\
$S_+, S_-$ & $+0.9\sigma$ & 0\% &
	$+0.2\sigma$ & 11\% \\
$K\pi, e^\pm$ & $-0.0\sigma$ & 11\% &
	$-0.1\sigma$ & 1\% \\
$S_+', e^\pm$ & $-0.3\sigma$ & 37\% &
	--- & $-19\%$ \\
$S_-, e^\pm$ & $-0.9\sigma$ & 54\% &
	--- & $-135\%$ \\
$K^0_L\pi^0, K\pi$ & --- & $-3\%$ &
	$-1.3\sigma$ & 14\% \\
$K^0_L\pi^0, S_-$ & --- & $-0\%$ &
	$+0.2\sigma$ & 5\% \\
\hline
\multicolumn{5}{c}{External or Other CLEO\nobreakdash-c (*) Inputs} \\
$R_{\rm WS}$, $R_{\rm M}$ & $-0.6\sigma$ & 3\% &
	--- & 100\% \\
${\cal B}(K^-\pi^+)$ & $-1.2\sigma$ & 1\% &
	--- & $-12\%$ \\
${\cal B}(K^+K^-)$ & $+1.4\sigma$ & 3\% &
	$+0.2\sigma$ & 29\% \\
${\cal B}(\pi^+\pi^-)$ & --- & $-0\%$ &
	--- & $-1\%$ \\
${\cal B}(K^0_S\pi^0)$ & $+0.4\sigma$ & 1\% &
	$-0.2\sigma$ & 3\% \\
${\cal B}(K^0_S\eta)$ & $+1.4\sigma$ & 0\% &
	$+0.0\sigma$ & $5\%$ \\
${\cal B}(K^0_S\omega)$ & $+0.1\sigma$ & 0\% &
	$-0.3\sigma$ & 0\% \\
${\cal B}(K^0_L\pi^0)$* & --- & $-5\%$ &
	$-1.2\sigma$ & 6\% \\
\hline\hline
\end{tabular}
\end{center}
\end{table}

\subsection{\boldmath Purity of $C$-Odd Initial State}\label{sec:initialPurity}

The purity of the $C$-odd $D^0\bar D^0$ initial state may be diluted by
a radiated photon, which would reverse the $C$ eigenvalue of the $D^0\bar D^0$
system and thus bias the fit results.  
ISR, FSR, and bremsstrahlung photon emission are benign because
they do not alter the relative angular momentum between the $D^0$ and
$\bar D^0$, occuring either before the $D^0\bar D^0$ are formed or after they
decay.  One problematic process would be photon radiation from the
$\psi(3770)$, resulting in a virtual $0^+$ state that then decays to
$D^0\bar D^0$.  This channel is suppressed, as there are no nearby $0^+$
states available.  Another possible contribution would be $\psi(3770)$ decay
to a virtual
$D^{*0}\bar D^0$, which subsequently decays to $D^0\bar D^0\gamma$.
Theoretical estimates of these amplitudes indicate a
$\psi(3770)\to D^0\bar D^0\gamma$
branching fraction of less than $10^{-8}$~\cite{petrov}.

\begin{figure}[htb]
\includegraphics*[width=\linewidth]{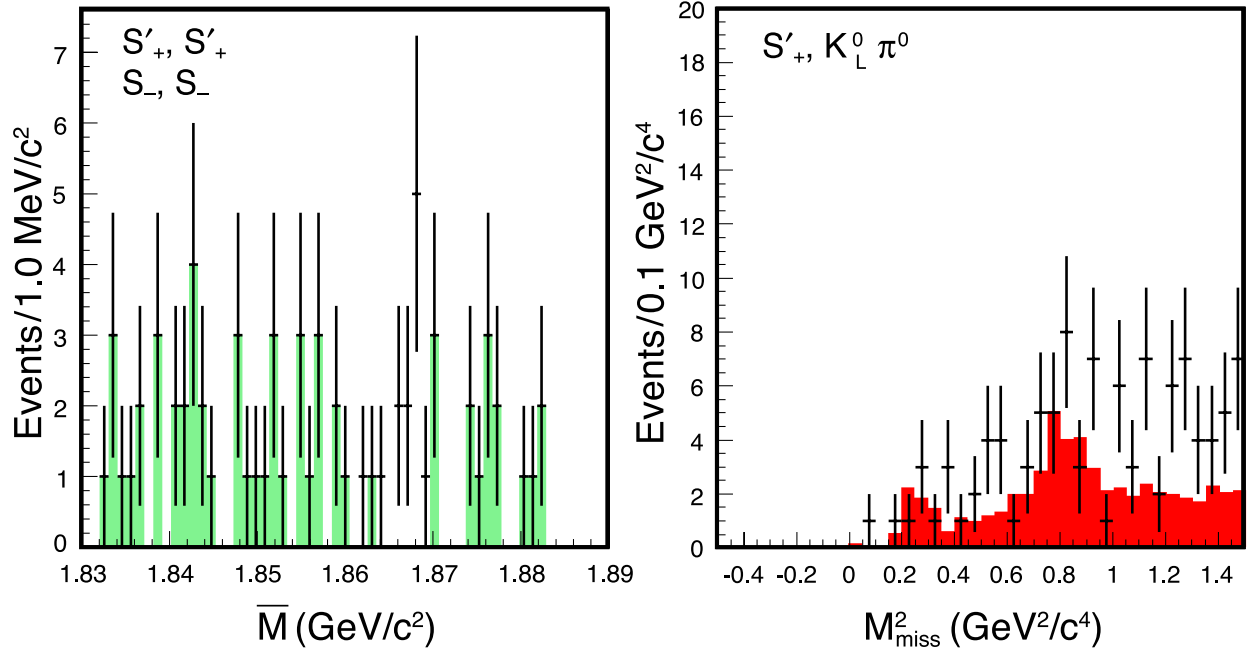}
\caption{Combined $\bar M$ distribution for fully reconstructed same-$CP$
$\{S_+', S_+'\}$ and $\{S_-, S_-\}$ DT modes and combined $M^2_{\rm miss}$
distribution for $\{S_+', K^0_L\pi^0\}$ modes.  Data are shown as points with
error bars.  For $\{S_+', S_+'\}$ and $\{S_-, S_-\}$, the shaded histogram
shows events outside the signal region.  For $\{S_+', K^0_L\pi^0\}$,
the shaded histogram represents simulations of the peaking
backgrounds $D^0\to\pi^0\pi^0$, $K^0_S\pi^0$,
$\eta\pi^0$, and $K^{*0}\pi^0$.
}
\label{fig:sameCPDTs}
\end{figure}

We verify the absence of this effect in data by searching for
same-$CP$ DT signals, which would be maximally enhanced for the
$C$-even configuration.  As shown in Fig.~\ref{fig:sameCPDTs} and
Table~\ref{tab:DTYieldEffsC+}, all such modes have yields consistent with
zero.  To determine the $C$-even fraction of the $D^0\bar D^0$ sample,
${\cal N}_{C+}/{\cal N}_{C-}$, we perform a variant of the standard and
extended fits that include the 15 same-$CP$ DT yields
and efficiencies shown in Table~\ref{tab:DTYieldEffsC+}.
We then express each yield as a sum of $C$-odd and $C$-even contributions:
$N_{ij} = {\cal N}_{C-}A_i^2 A_j^2 R_{C-} + {\cal N}_{C+}A_i^2 A_j^2 R_{C+}$,
where $R_{C-}$ and $R_{C+}$ are the functions of $y$, $r$, $\cos\delta$, and
$x\sin\delta$
given in Table~\ref{tab:C-evenRates}.  The results of these fits are shown
in Table~\ref{tab:C-evenFit}.  The $C$-even fraction is found to be
consistent with zero, with an uncertainty of 2.4\%, and shifts of the fit
parameters from the nominal fit results are negligible compared to the
systematic uncertainties already assigned.

\begin{table}[htbp]
\begin{center}
\caption{Same-$CP$ double tag yields and efficiencies included in the
$C$-even-allowed fit.}
\label{tab:DTYieldEffsC+}
\begin{tabular}{ccc}
\hline\hline
Mode & Yield & Efficiency (\%) \\
\hline
$K^+K^-, K^+K^-$ &
        $-2.1\pm 1.5\pm 3.1$ &
        $32.4\pm 1.7$ \\
$K^+K^-, \pi^+\pi^-$ &
        $0.9\pm 1.6\pm 1.1$ &
        $38.9\pm 2.2$ \\
$K^+K^-, K^0_S\pi^0\pi^0$ &
        $0.0\pm 1.0\pm 0.0$ &
        $6.6\pm 0.5$ \\
$\pi^+\pi^-, \pi^+\pi^-$ &
        $1.2\pm 2.8\pm 4.8$ &
        $44.9\pm 5.1$ \\
$\pi^+\pi^-, K^0_S\pi^0\pi^0$ &
        $1.0\pm 1.0\pm 0.0$ &
        $9.2\pm 1.0$ \\
$K^0_S\pi^0\pi^0, K^0_S\pi^0\pi^0$ &
        $0.0\pm 1.0\pm 0.0$ &
        $1.5\pm 0.2$ \\
$K^0_S\pi^0, K^0_S\pi^0$ &
        $0.0\pm 1.0\pm 0.0$ &
        $8.9\pm 0.5$ \\
$K^0_S\pi^0, K^0_S\eta$ &
        $0.0\pm 1.0\pm 0.0$ &
        $2.9\pm 0.3$ \\
$K^0_S\pi^0, K^0_S\omega$ &
        $1.0\pm 1.0\pm 0.0$ &
        $3.3\pm 0.2$ \\
$K^0_S\eta, K^0_S\eta$ &
        $0.0\pm 1.0\pm 0.0$ &
        $0.7\pm 0.4$ \\
$K^0_S\eta, K^0_S\omega$ &
        $0.0\pm 1.0\pm 0.0$ &
        $1.0\pm 0.2$ \\
$K^0_S\omega, K^0_S\omega$ &
        $1.0\pm 1.0\pm 0.0$ &
        $1.5\pm 0.2$ \\
\hline
$K^0_L\pi^0, K^+K^-$ &
        $1.7\pm 2.1\pm 0.0$ &
        $28.2\pm 0.1$ \\
$K^0_L\pi^0, \pi^+\pi^-$ &
        $4.3\pm 2.4\pm 0.0$ &
        $38.5\pm 0.1$ \\
$K^0_L\pi^0, K^0_S\pi^0\pi^0$ &
        $2.7\pm 2.3\pm 0.0$ &
        $6.4\pm 0.1$ \\
\hline\hline
\end{tabular}
\end{center}
\end{table}

\begin{table}[htbp]
\begin{center}
\caption{Comparison of $C$-odd and $C$-even yields, normalized by
${\cal N}A^2_i$ for ST modes $i$ and by ${\cal N}A^2_iA^2_j$
for DT modes $\{i, j\}$, to leading order in $x$, $y$, and $r^2$, with
$y'\equiv y\cos\delta-x\sin\delta$ and
$\tilde y\equiv y\cos\delta+x\sin\delta$.
Charge conjugate modes are implied.}
\label{tab:C-evenRates}
\begin{tabular}{ccc}
\hline\hline
Mode & $R_{C-}$ & $R_{C+}$ \\
\hline
$K^-\pi^+$ & $1+2ry\cos\delta+r^2$ & $1+2ry\cos\delta+r^2$ \\
$S_+$ & $2(1 - y)$ & $2(1 - y)$ \\
$S_-$ & $2(1 + y)$ & $2(1 + y)$ \\
\hline
$K^-\pi^+, K^-\pi^+$	& $R_{\rm M}$
	& $2 r(r + y')$ \\
$K^-\pi^+, K^+\pi^-$ & $1+2r^2(1-2\cos^2\delta)$
	& $1-2r^2(1-2\cos^2\delta)+4r\tilde y$ \\
$K^-\pi^+, S_+$ & $1 + 2r\cos\delta+r^2$
	& $\left[1 - 2r\cos\delta+r^2\right](1 - 2y)$ \\
$K^-\pi^+, S_-$ & $1 - 2r\cos\delta+r^2$
	& $\left[1 + 2r\cos\delta+r^2\right](1 + 2y)$ \\
$K^-\pi^+, e^-$ & 1 & $1+2r\tilde y$ \\
$S_+, S_+$ & 0 & $2(1 - 2y)$ \\
$S_-, S_-$ & 0 & $2(1 + 2y)$ \\
$S_+, S_-$ & 4 & 0 \\
$S_+, e^-$ & 1 & $1 - 2y$ \\
$S_-, e^-$ & 1 & $1 + 2y$ \\
\hline\hline
\end{tabular}
\end{center}
\end{table}

\begin{table}[htb]
\begin{center}
\caption{Results from the fits allowing for a $C$-even component in the
initial state, with inputs from Table~\ref{tab:DTYieldEffsC+}.
Uncertainties are statistical and systematic, respectively.  We also give the
shifts in the fit results with respect to the nominal fits.
}
\label{tab:C-evenFit}
\begin{tabular}{ccccc}
\hline\hline
Parameter & Standard, $C$-even Allowed & Shift &
	Extended, $C$-even Allowed & Shift \\
\hline
${\cal N}_{C-}$ $(10^6)$ &
	$1.044\pm 0.029\pm 0.017$ & $+0.002$ &
	$1.044\pm 0.030\pm 0.017$ & $+0.002$ \\
${\cal N}_{C+}/{\cal N}_{C-}$ $(10^{-3})$ &
	$-1.5\pm 23.0\pm 2.2$ & --- &
	$-1.4\pm 23.6\pm 1.9$ & --- \\
$y$ $(10^{-3})$ &
	$-46\pm 59\pm 16$ & $-0.2$ &
	$6.5\pm 0.2\pm 2.1$ & $-0.0$ \\
$r^2$ $(10^{-3})$ &
	$8.0\pm 6.8\pm 1.9$ & $+0.008$ &
	$3.44\pm 0.01\pm 0.09$ & $-0.000$ \\
$\cos\delta$ &
	$1.02\pm 0.19\pm 0.06$ & $-0.003$ &
	$1.10\pm 0.35\pm 0.07$ & $-0.005$ \\
$x^2$ $(10^{-3})$ &
	$-1.5\pm 4.1\pm 3.8$ & $-0.2$ &
	$0.06\pm 0.01\pm 0.05$ & $+0.01$ \\
$x\sin\delta$ $(10^{-3})$ &
	0 (fixed) & --- &
	$4.4\pm 2.4\pm 2.9$ & $-0.04$ \\
${\cal B}(K^-\pi^+)$ (\%) &
	$3.78\pm 0.05\pm 0.05$ & $-0.001$ &
	$3.78\pm 0.05\pm 0.05$ & $-0.001$ \\
${\cal B}(K^-K^+)$ $(10^{-3})$ &
	$3.87\pm 0.06\pm 0.06$ & $+0.001$ &
	$3.88\pm 0.06\pm 0.06$ & $+0.001$ \\
${\cal B}(\pi^-\pi^+)$ $(10^{-3})$ &
	$1.36\pm 0.02\pm 0.03$ & $-0.000$ &
	$1.36\pm 0.02\pm 0.03$ & $-0.000$ \\
${\cal B}(K^0_S\pi^0\pi^0)$ $(10^{-3})$ &
	$8.33\pm 0.45\pm 0.42$ & $-0.011$ &
	$8.34\pm 0.44\pm 0.42$ & $-0.011$ \\
${\cal B}(K^0_S\pi^0$) (\%) &
	$1.14\pm 0.03\pm 0.03$ & $-0.001$ &
	$1.14\pm 0.03\pm 0.03$ & $-0.001$ \\
${\cal B}(K^0_S\eta)$ $(10^{-3})$ &
	$4.42\pm 0.15\pm 0.28$  & $-0.002$ &
	$4.41\pm 0.15\pm 0.28$  & $-0.002$ \\
${\cal B}(K^0_S\omega)$ (\%) &
	$1.11\pm 0.04\pm 0.05$  & $-0.006$ &
	$1.11\pm 0.04\pm 0.05$  & $-0.006$ \\
${\cal B}(X^- e^+\nu_e)$ (\%) &
	$6.54\pm 0.17\pm 0.17$  & $-0.001$ &
	$6.59\pm 0.16\pm 0.16$  & $-0.001$ \\
${\cal B}(K^0_L\pi^0$) (\%) &
	$1.01\pm 0.03\pm 0.02$ & $+0.000$ &
	$1.01\pm 0.03\pm 0.02$ & $+0.000$ \\
\hline
$\chi^2_{\rm fit}$/ndof &
	34.1/59 & &
	59.3/71 & \\
\hline\hline
\end{tabular}
\end{center}
\end{table}


\section{Summary}\label{sec:summary}

We employ a double tagging technique with quantum-correlated $D^0\bar D^0$
decays at the $\psi(3770)$ to perform a first measurement of
$\cos\delta = 1.03^{+0.31}_{-0.17}\pm 0.06$, where the
uncertainties are statistical and systematic, respectively.
Within the physical region, we find $|\delta| < 75^\circ$ at the 95\%
confidence level.
By including external inputs on $y$ and $y'$ in the fit, we find an
alternative value of $\cos\delta = 1.10\pm 0.35\pm 0.07$, as well as
$x\sin\delta = (4.4^{+2.7}_{-1.8}\pm 2.9)\times 10^{-3}$
and $\delta = (22^{+11}_{-12}$$^{+9}_{-11})^\circ$.


\acknowledgments
We thank Alexey Petrov, William Lockman, Alan Schwartz, Bostjan Golob, and
Brian Petersen for helpful discussions.
We gratefully acknowledge the effort of the CESR staff 
in providing us with excellent luminosity and running conditions. 
D.~Cronin-Hennessy and A.~Ryd thank the A.P.~Sloan Foundation. 
This work was supported by the National Science Foundation, 
the U.S. Department of Energy, and 
the Natural Sciences and Engineering Research Council of Canada. 


\appendix

\section{Corrected Yields}
For use by future experiments, we provide our efficiency-corrected,
background-subtracted yields in Tables~\ref{tab:correctedYields1}
and~\ref{tab:correctedYields2}.
The correlation matrix, including statistical and systematic uncertainties,
for these yields appears in 
Tables~\ref{tab:correctedYieldCorrs1}--\ref{tab:correctedYieldCorrs6}.

\begin{table}[htbp]
\begin{center}
\caption{Efficiency-corrected, background-subtracted single tag and
double tag yields input to the fits.  Uncertainties are statistical and
systematic combined.}
\label{tab:correctedYields1}
\begin{tabular}{lc}
\hline\hline
Mode & Yield \\
\hline
$K^-\pi^+$ & $39159 \pm 743$ \\
$K^+\pi^-$ & $39326 \pm 742$ \\
$K^+K^-$ & $8279 \pm 208$ \\
$\pi^+\pi^-$ & $2877 \pm 145$ \\
$K^0_S\pi^0\pi^0$ & $18759 \pm 1731$ \\
$K^0_S\pi^0$ & $24923 \pm 1167$ \\
$K^0_S\eta$ & $10078 \pm 790$ \\
$K^0_S\omega$ & $23903 \pm 1581$ \\
$K^-\pi^+, K^-\pi^+$ & $3.1 \pm 3.9$ \\
$K^-\pi^+, K^+\pi^-$ & $1460 \pm 78$ \\
$K^-\pi^+, K^+K^-$ & $199 \pm 25$ \\
$K^-\pi^+, \pi^+\pi^-$ & $53 \pm 11$ \\
$K^-\pi^+, K^0_S\pi^0\pi^0$ & $393 \pm 79$ \\
$K^-\pi^+, K^0_S\pi^0$ & $472 \pm 55$ \\
$K^-\pi^+, K^0_S\eta$ & $131 \pm 49$ \\
$K^-\pi^+, K^0_S\omega$ & $337 \pm 76$ \\
$K^+\pi^-, K^+\pi^-$ & $2.6 \pm 3.2$ \\
$K^+\pi^-, K^+K^-$ & $149 \pm 21$ \\
$K^+\pi^-, \pi^+\pi^-$ & $53 \pm 11$ \\
$K^+\pi^-, K^0_S\pi^0\pi^0$ & $409 \pm 80$ \\
$K^+\pi^-, K^0_S\pi^0$ & $401 \pm 51$ \\
$K^+\pi^-, K^0_S\eta$ & $145 \pm 50$ \\
$K^+\pi^-, K^0_S\omega$ & $370 \pm 77$ \\
$K^+K^-, K^+K^-$ & $-6.5 \pm 10.6$ \\
$K^+K^-, \pi^+\pi^-$ & $2.3 \pm 5.0$ \\
$K^+K^-, K^0_S\pi^0\pi^0$ & $-0.3 \pm 15.0$ \\
$K^+K^-, K^0_S\pi^0$ & $224 \pm 38$ \\
$K^+K^-, K^0_S\eta$ & $97 \pm 38$ \\
$K^+K^-, K^0_S\omega$ & $285 \pm 67$ \\
$\pi^+\pi^-, \pi^+\pi^-$ & $2.7 \pm 12.4$ \\
$\pi^+\pi^-, K^0_S\pi^0\pi^0$ & $11 \pm 11$ \\
$\pi^+\pi^-, K^0_S\pi^0$ & $67 \pm 19$ \\
$\pi^+\pi^-, K^0_S\eta$ & $24 \pm 18$ \\
\hline\hline
\end{tabular}
\end{center}
\end{table}

\begin{table}[htbp]
\begin{center}
\caption{Efficiency-corrected, background-subtracted single tag and
double tag yields input to the fits.  Uncertainties are statistical and
systematic combined.}
\label{tab:correctedYields2}
\begin{tabular}{lc}
\hline\hline
Mode & Yield \\
\hline
$\pi^+\pi^-, K^0_S\omega$ & $67 \pm 27$ \\
$K^0_S\pi^0\pi^0, K^0_S\pi^0\pi^0$ & $-0.7 \pm 65.8$ \\
$K^0_S\pi^0\pi^0, K^0_S\pi^0$ & $395 \pm 121$ \\
$K^0_S\pi^0\pi^0, K^0_S\eta$ & $327 \pm 168$ \\
$K^0_S\pi^0\pi^0, K^0_S\omega$ & $265 \pm 159$ \\
$K^0_S\pi^0, K^0_S\pi^0$ & $-2.5 \pm 11.0$ \\
$K^0_S\pi^0, K^0_S\eta$ & $-1.0 \pm 33.5$ \\
$K^0_S\pi^0, K^0_S\omega$ & $5.7 \pm 31.7$ \\
$K^0_S\eta, K^0_S\eta$ & $0 \pm 141$ \\
$K^0_S\eta, K^0_S\omega$ & $-9.4 \pm 98.1$ \\
$K^0_S\omega, K^0_S\omega$ & $47 \pm 66$ \\
$X^+e^-\bar\nu_e, K^-\pi^+$ & $2488 \pm 131$ \\
$X^+e^-\bar\nu_e, K^-K^+$ & $319 \pm 62$ \\
$X^+e^-\bar\nu_e, \pi^-\pi^+$ & $97 \pm 28$ \\
$X^+e^-\bar\nu_e, K^0_S\pi^0\pi^0$ & $418 \pm 289$ \\
$X^+e^-\bar\nu_e, K^0_S\pi^0$ & $903 \pm 119$ \\
$X^+e^-\bar\nu_e, K^0_S\eta$ & $349 \pm 94$ \\
$X^+e^-\bar\nu_e, K^0_S\omega$ & $601 \pm 202$ \\
$X^-e^+\nu_e, K^+\pi^-$ & $2653 \pm 135$ \\
$X^-e^+\nu_e, K^-K^+$ & $261 \pm 60$ \\
$X^-e^+\nu_e, \pi^-\pi^+$ & $79 \pm 22$ \\
$X^-e^+\nu_e, K^0_S\pi^0\pi^0$ & $518 \pm 190$ \\
$X^-e^+\nu_e, K^0_S\pi^0$ & $876 \pm 103$ \\
$X^-e^+\nu_e, K^0_S\eta$ & $349 \pm 110$ \\
$X^-e^+\nu_e, K^0_S\omega$ & $570 \pm 200$ \\
$K^0_L\pi^0, K^-\pi^+$ & $516 \pm 47$ \\
$K^0_L\pi^0, K^+\pi^-$ & $485 \pm 46$ \\
$K^0_L\pi^0, K^+K^-$ & $-6.8 \pm 7.8$ \\
$K^0_L\pi^0, \pi^+\pi^-$ & $7.1 \pm 6.3$ \\
$K^0_L\pi^0, K^0_S\pi^0\pi^0$ & $10 \pm 37$ \\
$K^0_L\pi^0, K^0_S\pi^0$ & $565 \pm 75$ \\
$K^0_L\pi^0, K^0_S\eta$ & $154 \pm 59$ \\
$K^0_L\pi^0, K^0_S\omega$ & $508 \pm 105$ \\
\hline\hline
\end{tabular}
\end{center}
\end{table}

\begin{table}[htb]
\begin{footnotesize}
\begin{center}
\caption{Correlation coefficients (\%), with statistical and systematic uncertainties,
for the efficiency-corrected, background-subtracted
yields in Tables~\ref{tab:correctedYields1} and~\ref{tab:correctedYields2}.
Coefficients of 100\% are represented by a dash (---).}
\label{tab:correctedYieldCorrs1}
\begin{tabular}{l|ccccccccccccccccccccccc}
\hline\hline
& \rotatebox{90}{$K^-\pi^+$}
& \rotatebox{90}{$K^+\pi^-$}
& \rotatebox{90}{$K^+K^-$}
& \rotatebox{90}{$\pi^+\pi^-$}
& \rotatebox{90}{$K^0_S\pi^0\pi^0$}
& \rotatebox{90}{$K^0_S\pi^0$}
& \rotatebox{90}{$K^0_S\eta$}
& \rotatebox{90}{$K^0_S\omega$}
& \rotatebox{90}{$K^-\pi^+, K^-\pi^+$}
& \rotatebox{90}{$K^-\pi^+, K^+\pi^-$}
& \rotatebox{90}{$K^-\pi^+, K^+K^-$}
& \rotatebox{90}{$K^-\pi^+, \pi^+\pi^-$}
& \rotatebox{90}{$K^-\pi^+, K^0_S\pi^0\pi^0$}
& \rotatebox{90}{$K^-\pi^+, K^0_S\pi^0$}
& \rotatebox{90}{$K^-\pi^+, K^0_S\eta$}
& \rotatebox{90}{$K^-\pi^+, K^0_S\omega$}
& \rotatebox{90}{$K^+\pi^-, K^+\pi^-$}
& \rotatebox{90}{$K^+\pi^-, K^+K^-$}
& \rotatebox{90}{$K^+\pi^-, \pi^+\pi^-$}
& \rotatebox{90}{$K^+\pi^-, K^0_S\pi^0\pi^0$}
& \rotatebox{90}{$K^+\pi^-, K^0_S\pi^0$}
& \rotatebox{90}{$K^+\pi^-, K^0_S\eta$}
& \rotatebox{90}{$K^+\pi^-, K^0_S\omega$} \\
\hline
$K^-\pi^+$ & --- & $88$ & $56$ & $46$ & $3$ & $7$ & $4$ & $18$ & $0$ & $58$ & $21$ & $16$ & $9$ & $14$ & $6$ & $15$ & $-0$ & $22$ & $16$ & $8$ & $13$ & $5$ & $13$ \\
$K^+\pi^-$ & & --- & $56$ & $46$ & $3$ & $7$ & $4$ & $18$ & $-0$ & $58$ & $19$ & $15$ & $8$ & $12$ & $5$ & $14$ & $1$ & $23$ & $17$ & $9$ & $15$ & $6$ & $14$ \\
$K^+K^-$ & & & --- & $25$ & $2$ & $5$ & $3$ & $10$ & $0$ & $34$ & $24$ & $9$ & $5$ & $8$ & $3$ & $8$ & $0$ & $26$ & $9$ & $5$ & $9$ & $3$ & $8$ \\
$\pi^+\pi^-$ & & & & --- & $1$ & $2$ & $1$ & $8$ & $0$ & $20$ & $6$ & $15$ & $3$ & $4$ & $2$ & $6$ & $0$ & $7$ & $16$ & $3$ & $5$ & $2$ & $6$ \\
$K^0_S\pi^0\pi^0$ & & & & & --- & $82$ & $5$ & $63$ & $0$ & $2$ & $1$ & $1$ & $43$ & $28$ & $1$ & $22$ & $0$ & $1$ & $1$ & $42$ & $31$ & $1$ & $22$ \\
$K^0_S\pi^0$ & & & & & & --- & $10$ & $70$ & $0$ & $4$ & $1$ & $1$ & $38$ & $35$ & $3$ & $25$ & $0$ & $2$ & $1$ & $37$ & $38$ & $3$ & $24$ \\
$K^0_S\eta$ & & & & & & & --- & $8$ & $0$ & $2$ & $1$ & $1$ & $2$ & $4$ & $23$ & $3$ & $0$ & $1$ & $1$ & $2$ & $4$ & $22$ & $3$ \\
$K^0_S\omega$ & & & & & & & & --- & $0$ & $12$ & $4$ & $4$ & $30$ & $26$ & $3$ & $37$ & $0$ & $4$ & $4$ & $30$ & $28$ & $3$ & $36$ \\
$K^-\pi^+, K^-\pi^+$ & & & & & & & & & --- & $-2$ & $0$ & $0$ & $0$ & $0$ & $0$ & $0$ & $0$ & $0$ & $0$ & $0$ & $0$ & $0$ & $0$ \\
$K^-\pi^+, K^+\pi^-$ & & & & & & & & & & --- & $13$ & $11$ & $5$ & $8$ & $4$ & $9$ & $-2$ & $15$ & $11$ & $5$ & $9$ & $3$ & $9$ \\
$K^-\pi^+, K^+K^-$ & & & & & & & & & & & --- & $4$ & $2$ & $3$ & $1$ & $3$ & $0$ & $6$ & $4$ & $2$ & $3$ & $1$ & $3$ \\
$K^-\pi^+, \pi^+\pi^-$ & & & & & & & & & & & & --- & $2$ & $2$ & $1$ & $3$ & $0$ & $4$ & $4$ & $2$ & $3$ & $1$ & $3$ \\
$K^-\pi^+, K^0_S\pi^0\pi^0$ & & & & & & & & & & & & & --- & $14$ & $1$ & $11$ & $0$ & $2$ & $2$ & $19$ & $15$ & $1$ & $11$ \\
$K^-\pi^+, K^0_S\pi^0$ & & & & & & & & & & & & & & --- & $2$ & $10$ & $0$ & $3$ & $2$ & $14$ & $14$ & $2$ & $10$ \\
$K^-\pi^+, K^0_S\eta$ & & & & & & & & & & & & & & & --- & $2$ & $0$ & $1$ & $1$ & $1$ & $2$ & $5$ & $2$ \\
$K^-\pi^+, K^0_S\omega$ & & & & & & & & & & & & & & & & --- & $0$ & $4$ & $3$ & $11$ & $11$ & $2$ & $13$ \\
$K^+\pi^-, K^+\pi^-$ & & & & & & & & & & & & & & & & & --- & $0$ & $0$ & $0$ & $0$ & $0$ & $0$ \\
$K^+\pi^-, K^+K^-$ & & & & & & & & & & & & & & & & & & --- & $4$ & $2$ & $4$ & $1$ & $4$ \\
$K^+\pi^-, \pi^+\pi^-$ & & & & & & & & & & & & & & & & & & & --- & $2$ & $3$ & $1$ & $3$ \\
$K^+\pi^-, K^0_S\pi^0\pi^0$ & & & & & & & & & & & & & & & & & & & & --- & $15$ & $1$ & $11$ \\
$K^+\pi^-, K^0_S\pi^0$ & & & & & & & & & & & & & & & & & & & & & --- & $2$ & $11$ \\
$K^+\pi^-, K^0_S\eta$ & & & & & & & & & & & & & & & & & & & & & & --- & $1$ \\
$K^+\pi^-, K^0_S\omega$ & & & & & & & & & & & & & & & & & & & & & & & --- \\
\hline\hline
\end{tabular}
\end{center}
\end{footnotesize}
\end{table}

\begin{table}[htb]
\begin{footnotesize}
\begin{center}
\caption{Correlation coefficients (\%), with statistical and systematic uncertainties,
for the efficiency-corrected, background-subtracted
yields in Tables~\ref{tab:correctedYields1} and~\ref{tab:correctedYields2}.
}
\label{tab:correctedYieldCorrs2}
\begin{tabular}{l|ccccccccccccccccccccccc}
\hline\hline
& \rotatebox{90}{$K^+K^-, K^+K^-$}
& \rotatebox{90}{$K^+K^-, \pi^+\pi^-$}
& \rotatebox{90}{$K^+K^-, K^0_S\pi^0\pi^0$}
& \rotatebox{90}{$K^+K^-, K^0_S\pi^0$}
& \rotatebox{90}{$K^+K^-, K^0_S\eta$}
& \rotatebox{90}{$K^+K^-, K^0_S\omega$}
& \rotatebox{90}{$\pi^+\pi^-, \pi^+\pi^-$}
& \rotatebox{90}{$\pi^+\pi^-, K^0_S\pi^0\pi^0$}
& \rotatebox{90}{$\pi^+\pi^-, K^0_S\pi^0$}
& \rotatebox{90}{$\pi^+\pi^-, K^0_S\eta$}
& \rotatebox{90}{$\pi^+\pi^-, K^0_S\omega$}
& \rotatebox{90}{$K^0_S\pi^0\pi^0, K^0_S\pi^0\pi^0$}
& \rotatebox{90}{$K^0_S\pi^0\pi^0, K^0_S\pi^0$}
& \rotatebox{90}{$K^0_S\pi^0\pi^0, K^0_S\eta$}
& \rotatebox{90}{$K^0_S\pi^0\pi^0, K^0_S\omega$}
& \rotatebox{90}{$K^0_S\pi^0, K^0_S\pi^0$}
& \rotatebox{90}{$K^0_S\pi^0, K^0_S\eta$}
& \rotatebox{90}{$K^0_S\pi^0, K^0_S\omega$}
& \rotatebox{90}{$K^0_S\eta, K^0_S\eta$}
& \rotatebox{90}{$K^0_S\eta, K^0_S\omega$}
& \rotatebox{90}{$K^0_S\omega, K^0_S\omega$} \\
\hline
$K^-\pi^+$ & $-0$ & $-0$ & $0$ & $7$ & $3$ & $6$ & $-0$ & $0$ & $6$ & $2$ & $6$ & $0$ & $2$ & $0$ & $3$ & $0$ & $0$ & $1$ & $-0$ & $0$ & $1$ \\
$K^+\pi^-$ & $-0$ & $-0$ & $0$ & $7$ & $3$ & $6$ & $-0$ & $0$ & $6$ & $2$ & $6$ & $0$ & $2$ & $0$ & $3$ & $0$ & $0$ & $1$ & $-0$ & $0$ & $1$ \\
$K^+K^-$ & $6$ & $2$ & $1$ & $13$ & $5$ & $9$ & $-0$ & $0$ & $3$ & $1$ & $3$ & $0$ & $1$ & $0$ & $2$ & $0$ & $0$ & $1$ & $-0$ & $0$ & $0$ \\
$\pi^+\pi^-$ & $-0$ & $2$ & $0$ & $2$ & $1$ & $2$ & $11$ & $1$ & $8$ & $3$ & $7$ & $0$ & $1$ & $0$ & $2$ & $0$ & $0$ & $0$ & $-0$ & $0$ & $0$ \\
$K^0_S\pi^0\pi^0$ & $-0$ & $-0$ & $1$ & $19$ & $1$ & $11$ & $-0$ & $0$ & $13$ & $1$ & $10$ & $1$ & $42$ & $9$ & $33$ & $2$ & $0$ & $6$ & $-0$ & $0$ & $2$ \\
$K^0_S\pi^0$ & $-0$ & $-0$ & $0$ & $24$ & $2$ & $13$ & $-0$ & $0$ & $16$ & $1$ & $11$ & $0$ & $41$ & $8$ & $31$ & $2$ & $0$ & $7$ & $-0$ & $0$ & $2$ \\
$K^0_S\eta$ & $-0$ & $-0$ & $0$ & $3$ & $15$ & $2$ & $-0$ & $0$ & $2$ & $11$ & $1$ & $0$ & $3$ & $9$ & $3$ & $0$ & $1$ & $1$ & $2$ & $1$ & $0$ \\
$K^0_S\omega$ & $-0$ & $-0$ & $0$ & $17$ & $2$ & $17$ & $-0$ & $0$ & $13$ & $2$ & $14$ & $0$ & $31$ & $6$ & $30$ & $1$ & $0$ & $20$ & $-0$ & $3$ & $9$ \\
$K^-\pi^+, K^-\pi^+$ & $-0$ & $-0$ & $0$ & $0$ & $0$ & $0$ & $-0$ & $0$ & $0$ & $0$ & $0$ & $0$ & $0$ & $0$ & $0$ & $0$ & $0$ & $0$ & $-0$ & $0$ & $0$ \\
$K^-\pi^+, K^+\pi^-$ & $-0$ & $-0$ & $0$ & $5$ & $2$ & $4$ & $-0$ & $0$ & $4$ & $2$ & $4$ & $0$ & $1$ & $0$ & $2$ & $0$ & $0$ & $1$ & $-0$ & $0$ & $0$ \\
$K^-\pi^+, K^+K^-$ & $-0$ & $-0$ & $0$ & $2$ & $1$ & $2$ & $-0$ & $0$ & $1$ & $1$ & $1$ & $0$ & $0$ & $0$ & $1$ & $0$ & $0$ & $0$ & $-0$ & $0$ & $0$ \\
$K^-\pi^+, \pi^+\pi^-$ & $-0$ & $-0$ & $0$ & $1$ & $0$ & $1$ & $-0$ & $0$ & $2$ & $1$ & $2$ & $0$ & $0$ & $0$ & $1$ & $0$ & $0$ & $0$ & $-0$ & $0$ & $0$ \\
$K^-\pi^+, K^0_S\pi^0\pi^0$ & $-0$ & $-0$ & $0$ & $9$ & $1$ & $6$ & $-0$ & $0$ & $7$ & $0$ & $5$ & $0$ & $19$ & $4$ & $15$ & $1$ & $0$ & $3$ & $-0$ & $0$ & $1$ \\
$K^-\pi^+, K^0_S\pi^0$ & $-0$ & $-0$ & $0$ & $9$ & $1$ & $5$ & $-0$ & $0$ & $6$ & $1$ & $5$ & $0$ & $14$ & $3$ & $11$ & $1$ & $0$ & $2$ & $-0$ & $0$ & $1$ \\
$K^-\pi^+, K^0_S\eta$ & $-0$ & $-0$ & $0$ & $1$ & $3$ & $1$ & $-0$ & $0$ & $1$ & $3$ & $1$ & $0$ & $1$ & $2$ & $1$ & $0$ & $0$ & $0$ & $-0$ & $0$ & $0$ \\
$K^-\pi^+, K^0_S\omega$ & $-0$ & $-0$ & $0$ & $7$ & $1$ & $6$ & $-0$ & $0$ & $5$ & $1$ & $5$ & $0$ & $11$ & $2$ & $11$ & $0$ & $0$ & $8$ & $-0$ & $1$ & $3$ \\
$K^+\pi^-, K^+\pi^-$ & $-0$ & $-0$ & $0$ & $0$ & $0$ & $0$ & $-0$ & $0$ & $0$ & $0$ & $0$ & $0$ & $0$ & $0$ & $0$ & $0$ & $0$ & $0$ & $-0$ & $0$ & $0$ \\
$K^+\pi^-, K^+K^-$ & $-0$ & $-0$ & $0$ & $3$ & $1$ & $2$ & $-0$ & $0$ & $1$ & $1$ & $2$ & $0$ & $0$ & $0$ & $1$ & $0$ & $0$ & $0$ & $-0$ & $0$ & $0$ \\
$K^+\pi^-, \pi^+\pi^-$ & $-0$ & $-0$ & $0$ & $1$ & $1$ & $1$ & $-0$ & $0$ & $2$ & $1$ & $2$ & $0$ & $0$ & $0$ & $1$ & $0$ & $0$ & $0$ & $-0$ & $0$ & $0$ \\
$K^+\pi^-, K^0_S\pi^0\pi^0$ & $-0$ & $-0$ & $0$ & $9$ & $1$ & $6$ & $-0$ & $0$ & $7$ & $0$ & $5$ & $0$ & $19$ & $4$ & $15$ & $1$ & $0$ & $3$ & $-0$ & $0$ & $1$ \\
$K^+\pi^-, K^0_S\pi^0$ & $-0$ & $-0$ & $0$ & $9$ & $1$ & $6$ & $-0$ & $0$ & $7$ & $1$ & $5$ & $0$ & $15$ & $3$ & $12$ & $1$ & $0$ & $3$ & $-0$ & $0$ & $1$ \\
$K^+\pi^-, K^0_S\eta$ & $-0$ & $-0$ & $0$ & $1$ & $3$ & $1$ & $-0$ & $0$ & $1$ & $2$ & $1$ & $0$ & $1$ & $2$ & $1$ & $0$ & $0$ & $0$ & $-0$ & $0$ & $0$ \\
$K^+\pi^-, K^0_S\omega$ & $-0$ & $-0$ & $0$ & $7$ & $1$ & $6$ & $-0$ & $0$ & $5$ & $1$ & $5$ & $0$ & $11$ & $2$ & $10$ & $0$ & $0$ & $8$ & $-0$ & $1$ & $3$ \\
\hline\hline
\end{tabular}
\end{center}
\end{footnotesize}
\end{table}

\begin{table}[htb]
\begin{footnotesize}
\begin{center}
\caption{Correlation coefficients (\%), with statistical and systematic uncertainties,
for the efficiency-corrected, background-subtracted
yields in Tables~\ref{tab:correctedYields1} and~\ref{tab:correctedYields2}.
}
\label{tab:correctedYieldCorrs3}
\begin{tabular}{l|ccccccccccccccccccccccc}
\hline\hline
& \rotatebox{90}{$X^+e^-\bar\nu_e, K^-\pi^+$}
& \rotatebox{90}{$X^+e^-\bar\nu_e, K^-K^+$}
& \rotatebox{90}{$X^+e^-\bar\nu_e, \pi^-\pi^+$}
& \rotatebox{90}{$X^+e^-\bar\nu_e, K^0_S\pi^0\pi^0$}
& \rotatebox{90}{$X^+e^-\bar\nu_e, K^0_S\pi^0$}
& \rotatebox{90}{$X^+e^-\bar\nu_e, K^0_S\eta$}
& \rotatebox{90}{$X^+e^-\bar\nu_e, K^0_S\omega$}
& \rotatebox{90}{$X^-e^+\nu_e, K^+\pi^-$}
& \rotatebox{90}{$X^-e^+\nu_e, K^-K^+$}
& \rotatebox{90}{$X^-e^+\nu_e, \pi^-\pi^+$}
& \rotatebox{90}{$X^-e^+\nu_e, K^0_S\pi^0\pi^0$}
& \rotatebox{90}{$X^-e^+\nu_e, K^0_S\pi^0$}
& \rotatebox{90}{$X^-e^+\nu_e, K^0_S\eta$}
& \rotatebox{90}{$X^-e^+\nu_e, K^0_S\omega$}
& \rotatebox{90}{$K^0_L\pi^0, K^-\pi^+$}
& \rotatebox{90}{$K^0_L\pi^0, K^+\pi^-$}
& \rotatebox{90}{$K^0_L\pi^0, K^+K^-$}
& \rotatebox{90}{$K^0_L\pi^0, \pi^+\pi^-$}
& \rotatebox{90}{$K^0_L\pi^0, K^0_S\pi^0\pi^0$}
& \rotatebox{90}{$K^0_L\pi^0, K^0_S\pi^0$}
& \rotatebox{90}{$K^0_L\pi^0, K^0_S\eta$}
& \rotatebox{90}{$K^0_L\pi^0, K^0_S\omega$} \\
\hline
$K^-\pi^+$ & $43$ & $7$ & $6$ & $1$ & $4$ & $2$ & $6$ & $34$ & $7$ & $8$ & $2$ & $5$ & $2$ & $6$ & $17$ & $15$ & $-0$ & $-0$ & $-0$ & $2$ & $1$ & $5$ \\
$K^+\pi^-$ & $36$ & $7$ & $6$ & $1$ & $4$ & $2$ & $6$ & $43$ & $7$ & $8$ & $2$ & $5$ & $2$ & $6$ & $15$ & $18$ & $-0$ & $-0$ & $-0$ & $2$ & $1$ & $5$ \\
$K^+K^-$ & $22$ & $28$ & $3$ & $1$ & $3$ & $1$ & $3$ & $22$ & $27$ & $4$ & $1$ & $3$ & $1$ & $3$ & $9$ & $10$ & $0$ & $-0$ & $-0$ & $2$ & $1$ & $3$ \\
$\pi^+\pi^-$ & $14$ & $2$ & $18$ & $1$ & $2$ & $1$ & $3$ & $13$ & $2$ & $17$ & $1$ & $2$ & $1$ & $3$ & $5$ & $6$ & $-0$ & $2$ & $-0$ & $1$ & $0$ & $3$ \\
$K^0_S\pi^0\pi^0$ & $2$ & $0$ & $0$ & $28$ & $26$ & $1$ & $16$ & $2$ & $0$ & $0$ & $34$ & $30$ & $1$ & $16$ & $35$ & $36$ & $-1$ & $-0$ & $0$ & $49$ & $13$ & $36$ \\
$K^0_S\pi^0$ & $3$ & $1$ & $1$ & $16$ & $36$ & $3$ & $18$ & $3$ & $1$ & $1$ & $24$ & $40$ & $2$ & $18$ & $35$ & $36$ & $-2$ & $-1$ & $-1$ & $55$ & $14$ & $38$ \\
$K^0_S\eta$ & $2$ & $0$ & $0$ & $1$ & $3$ & $29$ & $2$ & $2$ & $0$ & $0$ & $2$ & $4$ & $27$ & $2$ & $0$ & $1$ & $-1$ & $-0$ & $-0$ & $3$ & $23$ & $2$ \\
$K^0_S\omega$ & $8$ & $2$ & $2$ & $12$ & $23$ & $3$ & $31$ & $8$ & $2$ & $2$ & $19$ & $27$ & $2$ & $32$ & $29$ & $30$ & $-1$ & $-0$ & $-1$ & $40$ & $11$ & $39$ \\
$K^-\pi^+, K^-\pi^+$ & $0$ & $0$ & $0$ & $0$ & $0$ & $0$ & $0$ & $0$ & $0$ & $0$ & $0$ & $0$ & $0$ & $0$ & $0$ & $0$ & $-0$ & $-0$ & $-0$ & $0$ & $0$ & $0$ \\
$K^-\pi^+, K^+\pi^-$ & $24$ & $5$ & $4$ & $1$ & $3$ & $1$ & $4$ & $23$ & $5$ & $5$ & $1$ & $3$ & $1$ & $4$ & $10$ & $10$ & $-0$ & $-0$ & $-0$ & $1$ & $1$ & $4$ \\
$K^-\pi^+, K^+K^-$ & $8$ & $2$ & $1$ & $0$ & $1$ & $0$ & $1$ & $8$ & $2$ & $2$ & $0$ & $1$ & $0$ & $1$ & $3$ & $4$ & $-0$ & $-0$ & $-0$ & $0$ & $0$ & $1$ \\
$K^-\pi^+, \pi^+\pi^-$ & $7$ & $1$ & $2$ & $0$ & $1$ & $0$ & $1$ & $7$ & $1$ & $3$ & $0$ & $1$ & $0$ & $1$ & $3$ & $3$ & $-0$ & $-0$ & $-0$ & $0$ & $0$ & $1$ \\
$K^-\pi^+, K^0_S\pi^0\pi^0$ & $4$ & $1$ & $1$ & $8$ & $12$ & $1$ & $8$ & $4$ & $1$ & $1$ & $12$ & $14$ & $1$ & $8$ & $17$ & $18$ & $-0$ & $-0$ & $-0$ & $23$ & $6$ & $17$ \\
$K^-\pi^+, K^0_S\pi^0$ & $6$ & $1$ & $1$ & $6$ & $11$ & $1$ & $7$ & $5$ & $1$ & $1$ & $9$ & $12$ & $1$ & $7$ & $8$ & $14$ & $-0$ & $-0$ & $-0$ & $18$ & $5$ & $14$ \\
$K^-\pi^+, K^0_S\eta$ & $2$ & $0$ & $0$ & $0$ & $1$ & $5$ & $1$ & $2$ & $0$ & $1$ & $0$ & $1$ & $5$ & $1$ & $1$ & $1$ & $-0$ & $-0$ & $-0$ & $1$ & $5$ & $1$ \\
$K^-\pi^+, K^0_S\omega$ & $6$ & $1$ & $1$ & $4$ & $8$ & $1$ & $9$ & $6$ & $1$ & $2$ & $7$ & $10$ & $1$ & $9$ & $12$ & $12$ & $-0$ & $-0$ & $-0$ & $14$ & $4$ & $13$ \\
$K^+\pi^-, K^+\pi^-$ & $0$ & $0$ & $0$ & $0$ & $0$ & $0$ & $0$ & $0$ & $0$ & $0$ & $0$ & $0$ & $0$ & $0$ & $0$ & $0$ & $-0$ & $-0$ & $-0$ & $0$ & $0$ & $0$ \\
$K^+\pi^-, K^+K^-$ & $10$ & $3$ & $2$ & $0$ & $1$ & $1$ & $1$ & $9$ & $3$ & $2$ & $0$ & $1$ & $0$ & $1$ & $4$ & $4$ & $-0$ & $-0$ & $-0$ & $1$ & $0$ & $1$ \\
$K^+\pi^-, \pi^+\pi^-$ & $7$ & $1$ & $2$ & $0$ & $1$ & $0$ & $1$ & $7$ & $1$ & $3$ & $0$ & $1$ & $0$ & $1$ & $3$ & $3$ & $-0$ & $-0$ & $-0$ & $0$ & $0$ & $1$ \\
$K^+\pi^-, K^0_S\pi^0\pi^0$ & $4$ & $1$ & $1$ & $8$ & $12$ & $1$ & $8$ & $4$ & $1$ & $1$ & $12$ & $14$ & $1$ & $8$ & $17$ & $18$ & $-0$ & $-0$ & $-0$ & $22$ & $6$ & $17$ \\
$K^+\pi^-, K^0_S\pi^0$ & $6$ & $1$ & $1$ & $6$ & $12$ & $1$ & $8$ & $6$ & $1$ & $1$ & $9$ & $14$ & $1$ & $8$ & $15$ & $10$ & $-0$ & $-0$ & $-0$ & $20$ & $5$ & $15$ \\
$K^+\pi^-, K^0_S\eta$ & $2$ & $0$ & $0$ & $0$ & $1$ & $5$ & $1$ & $2$ & $0$ & $1$ & $0$ & $1$ & $5$ & $1$ & $1$ & $1$ & $-0$ & $-0$ & $-0$ & $1$ & $5$ & $1$ \\
$K^+\pi^-, K^0_S\omega$ & $6$ & $1$ & $1$ & $4$ & $8$ & $1$ & $9$ & $6$ & $1$ & $1$ & $7$ & $9$ & $1$ & $9$ & $11$ & $12$ & $-0$ & $-0$ & $-0$ & $14$ & $4$ & $13$ \\
\hline\hline
\end{tabular}
\end{center}
\end{footnotesize}
\end{table}

\begin{table}[htb]
\begin{footnotesize}
\begin{center}
\caption{Correlation coefficients (\%), with statistical and systematic uncertainties,
for the efficiency-corrected, background-subtracted
yields in Tables~\ref{tab:correctedYields1} and~\ref{tab:correctedYields2}.
Coefficients of 100\% are represented by a dash (---).}
\label{tab:correctedYieldCorrs4}
\begin{tabular}{l|ccccccccccccccccccccccc}
\hline\hline
& \rotatebox{90}{$K^+K^-, K^+K^-$}
& \rotatebox{90}{$K^+K^-, \pi^+\pi^-$}
& \rotatebox{90}{$K^+K^-, K^0_S\pi^0\pi^0$}
& \rotatebox{90}{$K^+K^-, K^0_S\pi^0$}
& \rotatebox{90}{$K^+K^-, K^0_S\eta$}
& \rotatebox{90}{$K^+K^-, K^0_S\omega$}
& \rotatebox{90}{$\pi^+\pi^-, \pi^+\pi^-$}
& \rotatebox{90}{$\pi^+\pi^-, K^0_S\pi^0\pi^0$}
& \rotatebox{90}{$\pi^+\pi^-, K^0_S\pi^0$}
& \rotatebox{90}{$\pi^+\pi^-, K^0_S\eta$}
& \rotatebox{90}{$\pi^+\pi^-, K^0_S\omega$}
& \rotatebox{90}{$K^0_S\pi^0\pi^0, K^0_S\pi^0\pi^0$}
& \rotatebox{90}{$K^0_S\pi^0\pi^0, K^0_S\pi^0$}
& \rotatebox{90}{$K^0_S\pi^0\pi^0, K^0_S\eta$}
& \rotatebox{90}{$K^0_S\pi^0\pi^0, K^0_S\omega$}
& \rotatebox{90}{$K^0_S\pi^0, K^0_S\pi^0$}
& \rotatebox{90}{$K^0_S\pi^0, K^0_S\eta$}
& \rotatebox{90}{$K^0_S\pi^0, K^0_S\omega$}
& \rotatebox{90}{$K^0_S\eta, K^0_S\eta$}
& \rotatebox{90}{$K^0_S\eta, K^0_S\omega$}
& \rotatebox{90}{$K^0_S\omega, K^0_S\omega$} \\
\hline
$K^+K^-, K^+K^-$ & --- & $0$ & $-0$ & $-0$ & $-0$ & $-0$ & $0$ & $-0$ & $-0$ & $-0$ & $-0$ & $-0$ & $-0$ & $-0$ & $-0$ & $-0$ & $-0$ & $-0$ & $0$ & $-0$ & $-0$ \\
$K^+K^-, \pi^+\pi^-$ & & --- & $-0$ & $-0$ & $-0$ & $-0$ & $0$ & $0$ & $-0$ & $-0$ & $-0$ & $-0$ & $-0$ & $-0$ & $-0$ & $-0$ & $-0$ & $-0$ & $0$ & $-0$ & $-0$ \\
$K^+K^-, K^0_S\pi^0\pi^0$ & & & --- & $0$ & $0$ & $0$ & $-0$ & $0$ & $0$ & $0$ & $0$ & $0$ & $0$ & $0$ & $0$ & $0$ & $0$ & $0$ & $-0$ & $0$ & $0$ \\
$K^+K^-, K^0_S\pi^0$ & & & & --- & $1$ & $4$ & $-0$ & $0$ & $4$ & $0$ & $3$ & $0$ & $10$ & $2$ & $8$ & $0$ & $0$ & $2$ & $-0$ & $0$ & $1$ \\
$K^+K^-, K^0_S\eta$ & & & & & --- & $1$ & $-0$ & $0$ & $0$ & $1$ & $0$ & $0$ & $1$ & $1$ & $0$ & $0$ & $0$ & $0$ & $-0$ & $0$ & $0$ \\
$K^+K^-, K^0_S\omega$ & & & & & & --- & $-0$ & $0$ & $2$ & $0$ & $2$ & $0$ & $6$ & $1$ & $5$ & $0$ & $0$ & $2$ & $-0$ & $0$ & $1$ \\
$\pi^+\pi^-, \pi^+\pi^-$ & & & & & & & --- & $0$ & $-0$ & $-0$ & $-0$ & $-0$ & $-0$ & $-0$ & $-0$ & $-0$ & $-0$ & $-0$ & $0$ & $-0$ & $-0$ \\
$\pi^+\pi^-, K^0_S\pi^0\pi^0$ & & & & & & & & --- & $0$ & $0$ & $0$ & $0$ & $0$ & $0$ & $0$ & $0$ & $0$ & $0$ & $-0$ & $0$ & $0$ \\
$\pi^+\pi^-, K^0_S\pi^0$ & & & & & & & & & --- & $1$ & $3$ & $0$ & $7$ & $1$ & $5$ & $0$ & $0$ & $1$ & $-0$ & $0$ & $0$ \\
$\pi^+\pi^-, K^0_S\eta$ & & & & & & & & & & --- & $0$ & $0$ & $0$ & $1$ & $0$ & $0$ & $0$ & $0$ & $-0$ & $0$ & $0$ \\
$\pi^+\pi^-, K^0_S\omega$ & & & & & & & & & & & --- & $0$ & $5$ & $1$ & $4$ & $0$ & $0$ & $2$ & $-0$ & $0$ & $1$ \\
$K^0_S\pi^0\pi^0, K^0_S\pi^0\pi^0$ & & & & & & & & & & & & --- & $0$ & $0$ & $0$ & $0$ & $0$ & $0$ & $-0$ & $0$ & $0$ \\
$K^0_S\pi^0\pi^0, K^0_S\pi^0$ & & & & & & & & & & & & & --- & $4$ & $15$ & $1$ & $0$ & $3$ & $-0$ & $0$ & $1$ \\
$K^0_S\pi^0\pi^0, K^0_S\eta$ & & & & & & & & & & & & & & --- & $3$ & $0$ & $0$ & $1$ & $-0$ & $0$ & $0$ \\
$K^0_S\pi^0\pi^0, K^0_S\omega$ & & & & & & & & & & & & & & & --- & $1$ & $0$ & $4$ & $-0$ & $0$ & $2$ \\
$K^0_S\pi^0, K^0_S\pi^0$ & & & & & & & & & & & & & & & & --- & $0$ & $0$ & $-0$ & $0$ & $0$ \\
$K^0_S\pi^0, K^0_S\eta$ & & & & & & & & & & & & & & & & & --- & $0$ & $-0$ & $0$ & $0$ \\
$K^0_S\pi^0, K^0_S\omega$ & & & & & & & & & & & & & & & & & & --- & $-0$ & $1$ & $4$ \\
$K^0_S\eta, K^0_S\eta$ & & & & & & & & & & & & & & & & & & & --- & $-0$ & $-0$ \\
$K^0_S\eta, K^0_S\omega$ & & & & & & & & & & & & & & & & & & & & --- & $1$ \\
$K^0_S\omega, K^0_S\omega$ & & & & & & & & & & & & & & & & & & & & & --- \\
\hline\hline
\end{tabular}
\end{center}
\end{footnotesize}
\end{table}

\begin{table}[htb]
\begin{footnotesize}
\begin{center}
\caption{Correlation coefficients (\%), with statistical and systematic uncertainties,
for the efficiency-corrected, background-subtracted
yields in Tables~\ref{tab:correctedYields1} and~\ref{tab:correctedYields2}.
}
\label{tab:correctedYieldCorrs5}
\begin{tabular}{l|ccccccccccccccccccccccc}
\hline\hline
& \rotatebox{90}{$X^+e^-\bar\nu_e, K^-\pi^+$}
& \rotatebox{90}{$X^+e^-\bar\nu_e, K^-K^+$}
& \rotatebox{90}{$X^+e^-\bar\nu_e, \pi^-\pi^+$}
& \rotatebox{90}{$X^+e^-\bar\nu_e, K^0_S\pi^0\pi^0$}
& \rotatebox{90}{$X^+e^-\bar\nu_e, K^0_S\pi^0$}
& \rotatebox{90}{$X^+e^-\bar\nu_e, K^0_S\eta$}
& \rotatebox{90}{$X^+e^-\bar\nu_e, K^0_S\omega$}
& \rotatebox{90}{$X^-e^+\nu_e, K^+\pi^-$}
& \rotatebox{90}{$X^-e^+\nu_e, K^-K^+$}
& \rotatebox{90}{$X^-e^+\nu_e, \pi^-\pi^+$}
& \rotatebox{90}{$X^-e^+\nu_e, K^0_S\pi^0\pi^0$}
& \rotatebox{90}{$X^-e^+\nu_e, K^0_S\pi^0$}
& \rotatebox{90}{$X^-e^+\nu_e, K^0_S\eta$}
& \rotatebox{90}{$X^-e^+\nu_e, K^0_S\omega$}
& \rotatebox{90}{$K^0_L\pi^0, K^-\pi^+$}
& \rotatebox{90}{$K^0_L\pi^0, K^+\pi^-$}
& \rotatebox{90}{$K^0_L\pi^0, K^+K^-$}
& \rotatebox{90}{$K^0_L\pi^0, \pi^+\pi^-$}
& \rotatebox{90}{$K^0_L\pi^0, K^0_S\pi^0\pi^0$}
& \rotatebox{90}{$K^0_L\pi^0, K^0_S\pi^0$}
& \rotatebox{90}{$K^0_L\pi^0, K^0_S\eta$}
& \rotatebox{90}{$K^0_L\pi^0, K^0_S\omega$} \\
\hline
$K^+K^-, K^+K^-$ & $-0$ & $-0$ & $-0$ & $-0$ & $-0$ & $-0$ & $-0$ & $-0$ & $-0$ & $-0$ & $-0$ & $-0$ & $-0$ & $-0$ & $-0$ & $-0$ & $0$ & $0$ & $0$ & $-0$ & $-0$ & $-0$ \\
$K^+K^-, \pi^+\pi^-$ & $-0$ & $-0$ & $-0$ & $-0$ & $-0$ & $-0$ & $-0$ & $-0$ & $-0$ & $-0$ & $-0$ & $-0$ & $-0$ & $-0$ & $-0$ & $-0$ & $0$ & $0$ & $0$ & $-0$ & $-0$ & $-0$ \\
$K^+K^-, K^0_S\pi^0\pi^0$ & $0$ & $0$ & $0$ & $0$ & $0$ & $0$ & $0$ & $0$ & $0$ & $0$ & $0$ & $0$ & $0$ & $0$ & $0$ & $0$ & $-0$ & $-0$ & $-0$ & $0$ & $0$ & $0$ \\
$K^+K^-, K^0_S\pi^0$ & $3$ & $1$ & $0$ & $4$ & $7$ & $1$ & $5$ & $3$ & $1$ & $1$ & $6$ & $8$ & $1$ & $5$ & $9$ & $10$ & $-27$ & $-0$ & $-0$ & $13$ & $3$ & $9$ \\
$K^+K^-, K^0_S\eta$ & $1$ & $0$ & $0$ & $0$ & $1$ & $3$ & $0$ & $1$ & $0$ & $0$ & $0$ & $1$ & $3$ & $0$ & $0$ & $1$ & $-0$ & $-0$ & $-0$ & $1$ & $3$ & $1$ \\
$K^+K^-, K^0_S\omega$ & $3$ & $1$ & $1$ & $2$ & $4$ & $1$ & $4$ & $3$ & $1$ & $1$ & $3$ & $5$ & $0$ & $4$ & $6$ & $6$ & $-0$ & $-0$ & $-0$ & $7$ & $2$ & $6$ \\
$\pi^+\pi^-, \pi^+\pi^-$ & $-0$ & $-0$ & $-0$ & $-0$ & $-0$ & $-0$ & $-0$ & $-0$ & $-0$ & $-0$ & $-0$ & $-0$ & $-0$ & $-0$ & $-0$ & $-0$ & $0$ & $0$ & $0$ & $-0$ & $-0$ & $-0$ \\
$\pi^+\pi^-, K^0_S\pi^0\pi^0$ & $0$ & $0$ & $-0$ & $0$ & $0$ & $0$ & $0$ & $0$ & $0$ & $-0$ & $0$ & $0$ & $0$ & $0$ & $0$ & $0$ & $-0$ & $-0$ & $-0$ & $0$ & $0$ & $0$ \\
$\pi^+\pi^-, K^0_S\pi^0$ & $3$ & $0$ & $1$ & $3$ & $5$ & $1$ & $3$ & $3$ & $0$ & $1$ & $4$ & $6$ & $0$ & $3$ & $7$ & $7$ & $-0$ & $-18$ & $-0$ & $9$ & $2$ & $7$ \\
$\pi^+\pi^-, K^0_S\eta$ & $1$ & $0$ & $0$ & $0$ & $0$ & $2$ & $0$ & $1$ & $0$ & $1$ & $0$ & $1$ & $2$ & $0$ & $0$ & $0$ & $-0$ & $-0$ & $-0$ & $0$ & $2$ & $0$ \\
$\pi^+\pi^-, K^0_S\omega$ & $3$ & $0$ & $1$ & $2$ & $4$ & $0$ & $3$ & $3$ & $0$ & $1$ & $3$ & $4$ & $0$ & $3$ & $5$ & $5$ & $-0$ & $-0$ & $-0$ & $6$ & $2$ & $6$ \\
$K^0_S\pi^0\pi^0, K^0_S\pi^0\pi^0$ & $0$ & $0$ & $0$ & $0$ & $0$ & $0$ & $0$ & $0$ & $0$ & $0$ & $0$ & $0$ & $0$ & $0$ & $0$ & $0$ & $-0$ & $-0$ & $-0$ & $0$ & $0$ & $0$ \\
$K^0_S\pi^0\pi^0, K^0_S\pi^0$ & $1$ & $0$ & $0$ & $8$ & $13$ & $1$ & $8$ & $1$ & $0$ & $0$ & $12$ & $15$ & $1$ & $8$ & $17$ & $17$ & $-0$ & $-0$ & $-21$ & $24$ & $6$ & $17$ \\
$K^0_S\pi^0\pi^0, K^0_S\eta$ & $0$ & $0$ & $0$ & $2$ & $3$ & $2$ & $2$ & $0$ & $0$ & $0$ & $2$ & $3$ & $1$ & $2$ & $3$ & $3$ & $-0$ & $-0$ & $-0$ & $5$ & $3$ & $3$ \\
$K^0_S\pi^0\pi^0, K^0_S\omega$ & $2$ & $0$ & $0$ & $6$ & $10$ & $1$ & $7$ & $2$ & $0$ & $0$ & $9$ & $12$ & $1$ & $8$ & $13$ & $14$ & $-0$ & $-0$ & $-0$ & $19$ & $5$ & $15$ \\
$K^0_S\pi^0, K^0_S\pi^0$ & $0$ & $0$ & $0$ & $0$ & $1$ & $0$ & $0$ & $0$ & $0$ & $0$ & $0$ & $1$ & $0$ & $0$ & $1$ & $1$ & $-0$ & $-0$ & $-0$ & $1$ & $0$ & $1$ \\
$K^0_S\pi^0, K^0_S\eta$ & $0$ & $0$ & $0$ & $0$ & $0$ & $0$ & $0$ & $0$ & $0$ & $0$ & $0$ & $0$ & $0$ & $0$ & $0$ & $0$ & $-0$ & $-0$ & $-0$ & $0$ & $0$ & $0$ \\
$K^0_S\pi^0, K^0_S\omega$ & $0$ & $0$ & $0$ & $1$ & $2$ & $0$ & $5$ & $0$ & $0$ & $0$ & $2$ & $2$ & $0$ & $5$ & $3$ & $3$ & $-0$ & $-0$ & $-0$ & $4$ & $1$ & $5$ \\
$K^0_S\eta, K^0_S\eta$ & $-0$ & $-0$ & $-0$ & $-0$ & $-0$ & $-0$ & $-0$ & $-0$ & $-0$ & $-0$ & $-0$ & $-0$ & $-0$ & $-0$ & $-0$ & $-0$ & $0$ & $0$ & $0$ & $-0$ & $-0$ & $-0$ \\
$K^0_S\eta, K^0_S\omega$ & $0$ & $0$ & $0$ & $0$ & $0$ & $0$ & $1$ & $0$ & $0$ & $0$ & $0$ & $0$ & $0$ & $1$ & $0$ & $0$ & $-0$ & $-0$ & $-0$ & $0$ & $0$ & $1$ \\
$K^0_S\omega, K^0_S\omega$ & $0$ & $0$ & $0$ & $0$ & $1$ & $0$ & $2$ & $0$ & $0$ & $0$ & $1$ & $1$ & $0$ & $2$ & $1$ & $1$ & $-0$ & $-0$ & $-0$ & $1$ & $0$ & $2$ \\
\hline\hline
\end{tabular}
\end{center}
\end{footnotesize}
\end{table}

\begin{table}[htb]
\begin{footnotesize}
\begin{center}
\caption{Correlation coefficients (\%), with statistical and systematic uncertainties,
for the efficiency-corrected, background-subtracted
yields in Tables~\ref{tab:correctedYields1} and~\ref{tab:correctedYields2}.
Coefficients of 100\% are represented by a dash (---).}
\label{tab:correctedYieldCorrs6}
\begin{tabular}{l|ccccccccccccccccccccccc}
\hline\hline
& \rotatebox{90}{$X^+e^-\bar\nu_e, K^-\pi^+$}
& \rotatebox{90}{$X^+e^-\bar\nu_e, K^-K^+$}
& \rotatebox{90}{$X^+e^-\bar\nu_e, \pi^-\pi^+$}
& \rotatebox{90}{$X^+e^-\bar\nu_e, K^0_S\pi^0\pi^0$}
& \rotatebox{90}{$X^+e^-\bar\nu_e, K^0_S\pi^0$}
& \rotatebox{90}{$X^+e^-\bar\nu_e, K^0_S\eta$}
& \rotatebox{90}{$X^+e^-\bar\nu_e, K^0_S\omega$}
& \rotatebox{90}{$X^-e^+\nu_e, K^+\pi^-$}
& \rotatebox{90}{$X^-e^+\nu_e, K^-K^+$}
& \rotatebox{90}{$X^-e^+\nu_e, \pi^-\pi^+$}
& \rotatebox{90}{$X^-e^+\nu_e, K^0_S\pi^0\pi^0$}
& \rotatebox{90}{$X^-e^+\nu_e, K^0_S\pi^0$}
& \rotatebox{90}{$X^-e^+\nu_e, K^0_S\eta$}
& \rotatebox{90}{$X^-e^+\nu_e, K^0_S\omega$}
& \rotatebox{90}{$K^0_L\pi^0, K^-\pi^+$}
& \rotatebox{90}{$K^0_L\pi^0, K^+\pi^-$}
& \rotatebox{90}{$K^0_L\pi^0, K^+K^-$}
& \rotatebox{90}{$K^0_L\pi^0, \pi^+\pi^-$}
& \rotatebox{90}{$K^0_L\pi^0, K^0_S\pi^0\pi^0$}
& \rotatebox{90}{$K^0_L\pi^0, K^0_S\pi^0$}
& \rotatebox{90}{$K^0_L\pi^0, K^0_S\eta$}
& \rotatebox{90}{$K^0_L\pi^0, K^0_S\omega$} \\
\hline
$X^+e^-\bar\nu_e, K^-\pi^+$ & --- & $8$ & $6$ & $3$ & $9$ & $4$ & $7$ & $36$ & $8$ & $8$ & $4$ & $10$ & $4$ & $7$ & $7$ & $7$ & $-0$ & $-0$ & $-0$ & $1$ & $1$ & $3$ \\
$X^+e^-\bar\nu_e, K^-K^+$ & & --- & $1$ & $1$ & $2$ & $1$ & $1$ & $8$ & $2$ & $2$ & $1$ & $2$ & $1$ & $1$ & $1$ & $1$ & $-0$ & $-0$ & $-0$ & $0$ & $0$ & $0$ \\
$X^+e^-\bar\nu_e, \pi^-\pi^+$ & & & --- & $0$ & $2$ & $1$ & $1$ & $6$ & $1$ & $2$ & $1$ & $2$ & $1$ & $1$ & $1$ & $1$ & $-0$ & $-0$ & $-0$ & $0$ & $0$ & $1$ \\
$X^+e^-\bar\nu_e, K^0_S\pi^0\pi^0$ & & & & --- & $6$ & $1$ & $4$ & $3$ & $1$ & $1$ & $5$ & $7$ & $1$ & $4$ & $7$ & $7$ & $-0$ & $-0$ & $-0$ & $10$ & $2$ & $7$ \\
$X^+e^-\bar\nu_e, K^0_S\pi^0$ & & & & & --- & $2$ & $7$ & $9$ & $2$ & $2$ & $9$ & $14$ & $2$ & $7$ & $12$ & $12$ & $-0$ & $-0$ & $-0$ & $17$ & $5$ & $12$ \\
$X^+e^-\bar\nu_e, K^0_S\eta$ & & & & & & --- & $1$ & $4$ & $1$ & $1$ & $1$ & $2$ & $5$ & $1$ & $0$ & $0$ & $-0$ & $-0$ & $-0$ & $1$ & $6$ & $1$ \\
$X^+e^-\bar\nu_e, K^0_S\omega$ & & & & & & & --- & $7$ & $1$ & $2$ & $6$ & $8$ & $1$ & $7$ & $8$ & $8$ & $-0$ & $-0$ & $-0$ & $10$ & $3$ & $9$ \\
$X^-e^+\nu_e, K^+\pi^-$ & & & & & & & & --- & $8$ & $8$ & $4$ & $10$ & $4$ & $7$ & $6$ & $6$ & $-0$ & $-0$ & $-0$ & $1$ & $0$ & $2$ \\
$X^-e^+\nu_e, K^-K^+$ & & & & & & & & & --- & $2$ & $1$ & $2$ & $1$ & $1$ & $1$ & $1$ & $-0$ & $-0$ & $-0$ & $0$ & $0$ & $0$ \\
$X^-e^+\nu_e, \pi^-\pi^+$ & & & & & & & & & & --- & $1$ & $2$ & $1$ & $2$ & $1$ & $1$ & $-0$ & $-0$ & $-0$ & $0$ & $0$ & $1$ \\
$X^-e^+\nu_e, K^0_S\pi^0\pi^0$ & & & & & & & & & & & --- & $10$ & $1$ & $6$ & $11$ & $11$ & $-0$ & $-0$ & $-0$ & $15$ & $4$ & $11$ \\
$X^-e^+\nu_e, K^0_S\pi^0$ & & & & & & & & & & & & --- & $2$ & $9$ & $13$ & $14$ & $-0$ & $-0$ & $-0$ & $20$ & $5$ & $14$ \\
$X^-e^+\nu_e, K^0_S\eta$ & & & & & & & & & & & & & --- & $1$ & $0$ & $0$ & $-0$ & $-0$ & $-0$ & $1$ & $5$ & $1$ \\
$X^-e^+\nu_e, K^0_S\omega$ & & & & & & & & & & & & & & --- & $8$ & $8$ & $-0$ & $-0$ & $-0$ & $10$ & $3$ & $10$ \\
$K^0_L\pi^0, K^-\pi^+$ & & & & & & & & & & & & & & & --- & $25$ & $-0$ & $-0$ & $0$ & $26$ & $7$ & $19$ \\
$K^0_L\pi^0, K^+\pi^-$ & & & & & & & & & & & & & & & & --- & $-0$ & $-0$ & $0$ & $26$ & $7$ & $20$ \\
$K^0_L\pi^0, K^+K^-$ & & & & & & & & & & & & & & & & & --- & $0$ & $0$ & $-0$ & $-0$ & $-0$ \\
$K^0_L\pi^0, \pi^+\pi^-$ & & & & & & & & & & & & & & & & & & --- & $0$ & $-0$ & $-0$ & $-0$ \\
$K^0_L\pi^0, K^0_S\pi^0\pi^0$ & & & & & & & & & & & & & & & & & & & --- & $-0$ & $-0$ & $-0$ \\
$K^0_L\pi^0, K^0_S\pi^0$ & & & & & & & & & & & & & & & & & & & & --- & $9$ & $24$ \\
$K^0_L\pi^0, K^0_S\eta$ & & & & & & & & & & & & & & & & & & & & & --- & $7$ \\
$K^0_L\pi^0, K^0_S\omega$ & & & & & & & & & & & & & & & & & & & & & & --- \\
\hline\hline
\end{tabular}
\end{center}
\end{footnotesize}
\end{table}




\end{document}